\documentclass[pre,preprint]{revtex4-1}

\usepackage{amsmath}  
\usepackage{amsfonts} 
\usepackage{graphicx} 
\usepackage{subfig}
\usepackage{enumitem} 
\usepackage{xcolor} 
\usepackage[utf8]{inputenc} 

\DeclareFontFamily{U}{BOONDOX-calo}{\skewchar\font=45 }
\DeclareFontShape{U}{BOONDOX-calo}{m}{n}{
  <-> s*[1.05] BOONDOX-r-calo}{}
\DeclareFontShape{U}{BOONDOX-calo}{b}{n}{
  <-> s*[1.05] BOONDOX-b-calo}{}
\DeclareMathAlphabet{\mathcalboondox}{U}{BOONDOX-calo}{m}{n}
\SetMathAlphabet{\mathcalboondox}{bold}{U}{BOONDOX-calo}{b}{n}
\DeclareMathAlphabet{\mathbcalboondox}{U}{BOONDOX-calo}{b}{n}

\begin{document}

\title{Fractional dynamics on circulant multiplex networks: optimal coupling and long-range navigation for continuous-time random walks}

\author{Alfonso Allen-Perkins}
\email{alfonso.allen.perkins@gmail.com}
\affiliation{Instituto de F\'isica, Universidade Federal da Bahia, 40210-210 Salvador, Brazil.}
\affiliation{Complex System Group, Universidad Polit\'ecnica de Madrid, 28040-Madrid, Spain.}

\author{Roberto F. S. Andrade}
\email{randrade@ufba.br}
\affiliation{Instituto de F\'isica, Universidade Federal da Bahia, 40210-210 Salvador, Brazil.}%
\date{\today}

\begin{abstract}
This work analyzes fractional continuous-time random walks on two-layer multiplexes. A node-centric dynamics is used, in which it is assumed a Poisson distribution of a walker to become active, while a jump to one of its neighbors depends on the connection weight. Synthetic multiplexes with well known topology are used to illustrate dynamical features obtained by numerical simulations, while exact analytical expressions are presented for multiplexes assembled by circulant layers with finite number of nodes. Special attention is given to the effect of inter- $D_x$ and intra-layer $D_i$ coefficients on the system's behavior. In opposition to usual discrete time dynamics, the relaxation time has a well defined minimum at an optimal value of $D_x/D_i$. It is found that, even for the enhanced diffusion condition, the walkers mean square displacement increases linearly with time.

\vspace{0.25cm}

\end{abstract}

\begin{titlepage}
\maketitle
\setcounter{page}{1}
\end{titlepage}

\section{Introduction}

Over the last two decades, modeling complex systems as networks has proven to be a sucessful approach to characterize the structure and dynamics of many real-world systems \cite{albert02,newman03,bocaletti06,barrat08}. Different dynamics have been investigated on top of networks, such as spreading processes \cite{arruda18}, percolation \cite{Dorogovtsev08}, or synchronization \cite{tang14,rodrigues16}.  Among these dynamical processes, diffusion and random walks (RWs) have been analyzed thoroughly. Indeed, RWs models have widespread use both in the analysis of diffusion and navigability in networks as in exploring their fine-grained organization \cite{noh04,klafter2011first,masuda17}. Most of the research on RWs relies on the nearest-neighbor (\textbf{NN}) paradigm \cite{noh04} in which the walker can only hop to one of the \textbf{NN}s of its current node (or position). However, other RWs definitions in discrete and continuous times, allow for both \textbf{NN} and long distance hops. Well known examples are, e.g., L\'evy RWs \cite{riascos12,guo16,estrada17b,nigris17}, RWs based on the $d$-path Laplacian operator \cite{estrada17b}, and those defined by fractional transport in networks \cite{riascos14,riascos15,michelitsch19}. Such non-\textbf{NN} strategies often correspond to the better options for randomly reaching a target in an unknown environment as, for instance, the foraging of species in a given environment \cite{lomholt2008levy,humphries2010environmental,song2010modelling,rhee2011levy,ghandi2011foraging}.

More recently, multiplex networks and other multilayer structures were identified as more comprehensive frameworks to describe those complex systems in which agents may interact through several channels. Here, links representing channels with different meaning and relevance are embedded in distinct layers \cite{boccaletti14,kivela14,battiston17,bianconi18}. As before, multiplexes have widespread use to describe, among others, social \cite{szell10,cozzo13,li15,arruda17}, biochemical \cite{cozzo12,battiston17}, or transportation systems \cite{dedomenico14,aleta17}. Here, layers may represent underground, bus and railway networks in large cities, each one  associated to a different spatial and temporal scale. A diffusion process in a transport system like this occurs both within and across layers \cite{arruda18,gomez13,cencetti19,dedomenico16,tejedor18}, accounting for the actual displacement to reach different places as within the change of embarking platforms. Multiplexes have also been used to study related dynamical systems, such as reaction-diffusion \cite{asllani14,kouvaris15,busiello18} and synchronization processes \cite{gambuzza15,sevilla15,genio16,allen17}. The eigenvalue spectrum of the supra-Laplacian matrix associated to the multiplex plays a key role in the description such diffusive process. Several results indicate the presence of super-diffusive behavior in undirected and directed multiplexes \cite{gomez13,sole13,radicchi13,cozzo12,cozzo16,sanchez14,cozzob16,arruda18b,serrano17,tejedor18,cencetti19}, meaning that their relaxation time to the steady state is smaller than any other observed for the isolated layer.

On the other hand, it seems that fractional diffusion on multiplex networks has not yet received due attention. In this work, we address this matter, following the framework introduced in Refs.~\cite{riascos14,riascos15,michelitsch19}. Our study considers the fractional diffusion of node-centric continuous-time RWs on undirected multiplex with two layers. We present numerical results of main dynamical features for several multiplexes with well known topology, and derive exact analytical expressions for circulant layers. An important aspect is the role played by the ratio between the inter- and intra-layer coefficients. Our results indicate a nonmonotonic behavior in the rate of convergence to the steady state as the inter-layer coefficient increases. The walkers mean square displacement (MSD$\equiv \left \langle r^2(t) \right \rangle$) illustrates the existence of an optimal diffusive regime depending on both the inter-layer coupling and on the fractional parameter. Following the nomenclature in \cite{dedomenico14}, we show that fractional dynamics turns \textit{classic random walkers} into a new type of \textit{physical random walkers}, which are allowed to (i) switch layer and (ii) perform long hops to another distant vertex in the same jump. In spite of such enhanced diffusion, our results also show that the MSD still increases linearly with time when the number of multiplex nodes is finite.

The paper is organized as follows. In Sec.~\ref{Sec:Diff}, we describe the diffusion dynamics on undirected multiplex networks and MSD. Section \ref{Sec:FracDiff} defines fractional dynamics on such systems. Analytical results for fractional random walks with continuous time on regular multiplex with circulant layers are presented in Sec.~\ref{Sec:results}. Finally, our conclusions are summarized in Sec.~\ref{conclusions}.

\section{Diffusion dynamics on multiplex networks}
\label{Sec:Diff}

Let us consider a multiplex $\mathcal{M}$ with $N$ nodes and $M$ layers. Let $\mathbf{A}^\alpha =\left ( \mathbf{A}_{ij}^\alpha \right )$ denote the adjacency matrix for the $\alpha$th layer with $1\leq \alpha \leq M$. In this work we focus on multiplexes $\mathcal{M}$ whose layers are undirected and unsigned (i.e., the edge weights are nonnegative), and contain no self-loops, i.e., $\mathbf{A}_{ij}^\alpha=\mathbf{A}_{ji}^\alpha=w_{ij}>0$ if there is a link between the nodes $i$ and $j$ in the layer $\alpha$ (and $i\neq j$), and 0 otherwise. If the layers of $\mathcal{M}$ are also unweighted, then $w_{ij}=1$. On the other hand, the strength $s_{i}^{\alpha}$ of a vertex $i$ with respect to its connections with other vertices $j$ (with $j=1,\cdots,N$) in the same layer $\alpha$ is given by $s_{i}^{\alpha}=\sum_{j=1}^N  \mathbf{A}_{ij}^\alpha$.

On a discrete space, diffusive phenomena are described in terms of Laplacian matrices, which can be formally obtained as a discretized version of the Laplacian operator $\left ( -\bigtriangledown ^2 \right )$ on regular lattices, and have been generalized for more complex topologies \cite{riascos14,riascos15}. In the case of multiplexes, we let $\vec{x}$ be a $NM\times1$ state (column) vector whose entry $i+(\alpha-1)N$ (with $i=1,\cdots,N$) describes the concentration of a generic flowing quantity at time $t$ on node $i$ at the $\alpha$th layer, $x_i^\alpha$. Therefore, the usual diffusion equation in matrix form reads:


\begin{equation}
\frac{\mathrm{d}\vec{x}(t) }{\mathrm{d} t}=-\mathbf{\mathcal{L}}^\mathcal{M}\vec{x}(t)
\label{sol_prob}
\end{equation}

\noindent where

\begin{equation}
\mathbf{\mathcal{L}}^\mathcal{M}=\mathbf{\mathcal{L}}^\ell+\mathbf{\mathcal{L}}^x
\label{def_comb_lap}
\end{equation}

\noindent denotes the $NM\times NM$ (combinatorial) supra-Laplacian matrix defined in \cite{masuda17,gomez13,cencetti19,sole13, bianconi18}, and $\mathcal{L}^\ell$ and $\mathcal{L}^x$ represent the intra-layer and the inter-layer supra-Laplacian matrices, respectively, given by:

\begin{equation}
\mathcal{L}^\ell=\left ( \begin{matrix}
D_1 \mathbf{L}^1 &  &  & \\
 & D_2 \mathbf{L}^2 &  & \\
 &  & \ddots  & \\
 &  &  & D_M \mathbf{L}^M
\end{matrix} \right ),
\end{equation}

and

\begin{equation}
\mathcal{L}^x=\left ( \begin{matrix}
\sum_\alpha D_{1 \alpha}\mathbf{I}_N & -D_{1 2}\mathbf{I}_N & \cdots & -D_{1 M}\mathbf{I}_N\\
-D_{21}\mathbf{I}_N & \sum_\alpha D_{2 \alpha}\mathbf{I}_N & \cdots & -D_{2M}\mathbf{I}_N\\
\vdots  & \vdots & \ddots  & \vdots \\
-D_{M 1}\mathbf{I}_N & -D_{M 2}\mathbf{I}_N & \cdots & \sum_\alpha D_{M \alpha}\mathbf{I}_N
\end{matrix} \right ).
\label{interlayer_connect_matrix}
\end{equation}

\noindent In the above equations, $\mathcal{L}^\ell$ is a block-diagonal matrix, $D_\alpha$ denotes the intra-layer diffusion constant in the $\alpha$th layer, $D_{\alpha\beta}$ (with $\alpha,\beta\in\left \{ 1,\cdots,M \right \}$ and $\beta\neq \alpha$) refers to the inter-layer diffusion constant between the $\alpha$th and $\beta$th layers, $\mathbf{I}_N$ represents the $N\times N$ identity matrix, and $\mathbf{L}^\alpha$ is the usual $N\times N$ (combinatorial) Laplacian matrix of the layer $\alpha$, with elements $\mathbf{L}^{\alpha}_{ij}=s_{i}^{\alpha}\delta_{ij} -  \mathbf{A}_{ij}^{\alpha}$, and $\delta$ is the Kronecker delta function. Thus, the matrix $\mathbf{\mathcal{L}}^\mathcal{M}$ represents the generalization of the graph Laplacian to the case of linear diffusion on multiplex networks. For simplicity we will consider only diffusion processes where $D_{\alpha\beta}=D_{\beta\alpha}$, so that $\mathcal{L}^x$ and  $\mathbf{\mathcal{L}}^\mathcal{M}$ are symmetric matrices.

Finally, according to Eq.~(\ref{def_comb_lap}), the elements of the main diagonal of $\mathbf{\mathcal{L}}^\mathcal{M}$ represent the total strength of a given node at a given layer, i.e., the sum of (i) the strength of such vertex with respect to its connections with other vertices in the same layer and (ii) the strength of the same vertex with respect to connections to its counterparts in different layers. To denote the total strength of node $i$ in layer $\alpha$, we introduce the following short-hand notation:

\begin{equation}
\sigma_i^\alpha=\left ( \mathbf{\mathcal{L}}^\mathcal{M} \right )_{ff},
\label{total_strenght}
\end{equation}

\noindent where $f=i+(\alpha-1)N$.


\subsection{CTRW\lowercase{s} and MSD on multiplex networks}

The usual \textit{discrete-time random walk} (DTRW) is a random sequence of vertices generated as follows: given a starting vertex $i$, denoted as ``origin of the walk'', at each discrete time step $t$, the walker jumps to one \textbf{NN} of its current node \cite{masuda17,aldous02,lovasz93,noh04}. In the case of multiplex networks, because of its peculiar interconnected structure, DTRW can also move from one layer to another one, provided that such layers ($\alpha$ and $\beta$) are connected with each other (i.e., $D_{\alpha\beta}\neq0$).

In the case of \textit{continuous-time random walks} (CTRWs), it is assumed that the duration of the walkers waiting times between two moves obeys a given probability density function \cite{masuda17}. For that reason, the actual timing of the moves must be taken into account. For the sake of simplicity, in this work we consider that the waiting times are distributed according to a Poisson distribution with constant rate (i.e.,the exponential distribution).

Here, it becomes necessary to distinguish between two different cases of Poissonian CTRWs: Node-centric and edge-centric RWs. The Poissonian node-centric CTRWs follow the same assumption of DTRWs: when a walker becomes active, it moves from its current node to one of the neighbors with a probability proportional to the weight of the connection between such nodes. On the other hand, in the Poissonian edge-centric CTRWs, each edge (rather than a node) is activated independently according to a renewal process. Thus, if a trajectory includes many nodes with large strengths, the number $n$ of moves in the time interval $[0,t]$ tends to be larger than for trajectories that traverse many nodes with small strengths. For a wider description of the specific features of each random walk the reader is referred to Ref.~\cite{masuda17}.

To generalize the fractional diffusion framework introduced in Refs.~\cite{riascos14,riascos15,michelitsch19} to multiplex networks, we restrict our analyzes to CTRWs. Let $\vec{p}$ be a $NM\times1$ vector whose entry $i+(\alpha-1)N$ (with $i=1,\cdots,N$) is the probability of finding the random walker at time $t$ on node $i$ at the $\alpha$th layer. The transition rules governing the diffusion dynamics of the node-centric random walks are determined a master equation which, in terms of suitably defined matrices, can be written as

\begin{equation}
\frac{\mathrm{d}\vec{p}(t)^T }{\mathrm{d} t}=-\vec{p}(t)^T\mathbf{\mathcal{S}}^{-1}\mathbf{\mathcal{L}}^\mathcal{M}=-\vec{p}(t)^T \mathcalboondox{L},
\label{prob_node_centric}
\end{equation}

\noindent  On the other hand, the dynamics of the edge-centric ones are described by

\begin{equation}
\frac{\mathrm{d}\vec{p}(t)^T }{\mathrm{d} t}=-\vec{p}(t)^T\mathbf{\mathcal{L}}^\mathcal{M}.
\label{prob_edge_centric}
\end{equation}

\noindent In Eqs.~(\ref{prob_node_centric}) and (\ref{prob_edge_centric}),  $\mathbf{X}^T$ stands for the transpose of matrix $\mathbf{X}$, and $\mathbf{\mathcal{S}}$ is the $NM\times NM$ diagonal matrix with elements $\mathbf{\mathcal{S}}_{ii}=\mathbf{\mathcal{L}}^\mathcal{M}_{ii}$. The $NM\times NM$ matrix $\mathcalboondox{L}$ denotes the ``random walk normalized supra-Laplacian" \cite{masuda17} (or just ``normalized supra-Laplacian" \cite{dedomenico14}). According to the definition of $\mathcalboondox{L}$, its elements can be expressed as,

\begin{equation}
\mathcalboondox{L}_{fg}=\frac{\mathbf{\mathcal{L}}^\mathcal{M}_{fg}}{\mathbf{\mathcal{L}}^\mathcal{M}_{ff}}=\delta_{fg}-\mathcalboondox{T}_{fg},
\label{transition_matrix}
\end{equation}

\noindent where $\mathcalboondox{T}_{fg}$ are the elements of the $NM\times NM$ transition matrix $\mathcalboondox{T}$ of a discrete-time random walk, describing transition probability from one node to its \textbf{NN}s in the corresponding layer or to the node's counterparts in different layers, with equal probability \cite{noh04,klafter2011first,riascos14,riascos15,michelitsch19,masuda17}. Indeed, note that $\mathcalboondox{T}$ is a stochastic matrix, that satisfies $\mathcalboondox{T}_{ff}=0$ and $\sum_{g=1}^{NM} \mathcalboondox{T}_{fg}=1$. Following \cite{dedomenico14,riascos14,riascos15,michelitsch19}, heareafter we will consider only the case of node-centric CTRW (or \textit{classical random walker} as in \cite{dedomenico14}).

The MSD, defined by $\left \langle r^2(t) \right \rangle$, is a measure of the ensemble average distance between the position of a walker at a time $t$, $x(t)$, and a reference position, $x_0$. Assuming that $\left \langle r^2(t) \right \rangle$ has a power law dependence with respect to time, we have

\begin{equation}
\mathrm{MSD}\equiv \left \langle r^2(t) \right \rangle = \left \langle \left ( x(t)-x_0 \right )^2 \right \rangle \sim  t^\varepsilon,
\end{equation}

\noindent where the value of the parameter $\varepsilon$ classifies the type of diffusion into normal diffusion ($\varepsilon=1$), sub-diffusion  ($\varepsilon<1$), or super-diffusion ($\varepsilon>1$). Although MSD is one of the used measures to analyze general stochastic data \cite{almaas03,gallos04}, in order to better characterize diffusion, additional measures are also required, e.g., first passage observables \cite{masuda17}. For the type of results we discuss here, $\left \langle r^2(t) \right \rangle$ is essential to provide a clear cut way to characterize the time dependence.

According to Eq.~(\ref{prob_node_centric}), the probability of finding the random walker at node $j$ in the $\beta$th layer (at time $t$), when the random walker was initially located at node $i$ in the $\alpha$th layer, is given by:

\begin{equation}
p(t)_{i^\alpha\rightarrow j^\beta}=\vec{p}(t)_g= \vec{\mathcal{C}}_g\,^T \exp\left (-t\mathcalboondox{L}\right )\vec{\mathcal{C}}_f,
\label{prob_walker_normal}
\end{equation}

\noindent where $g=j+(\beta-1)N$ and $f=i+(\alpha-1)N$ (with $i,j \in \left \{ 1, \cdots, N \right \}$ and $\alpha,\beta \in \left \{ 1, \cdots, M \right \}$), and $\vec{\mathcal{C}}_\ell$ represents the ($NM\times1$) $\ell-$th vector of the canonical base of $\mathbb{R}^{NM}$ with components $\left ( \vec{\mathcal{C}}_g \right )_m=\delta_{m\ell}$). Therefore, in the case of node-centric CTRWs, we can quantify $\left \langle r^2(t) \right \rangle$ at time $t$ as follows:

\begin{equation}
\left \langle r^2(t) \right \rangle = \frac{1}{N^2M^2}\sum_{\alpha=1}^M \sum_{\beta=1}^{M} \sum_{i=1}^N \sum_{j=1}^{N}\left ( d_{i^\alpha\rightarrow j^\beta} \right )^2 p(t)_{i^\alpha\rightarrow j^\beta},
\label{eq_MSD_multpl_new}
\end{equation}

\noindent where $d_{i^\alpha\rightarrow j^\beta}$ is the length of the shortest path distance between $i$ in the $\alpha$th layer and $j$ in the $\beta$th layer, that is, the smallest number of edges connecting those nodes.

\section{Fractional diffusion on multiplex networks}
\label{Sec:FracDiff}

\subsection{General Case}
\label{Sec:FracDiff_general}

In this sub-section we present the general expressions for the combinatorial and normalized supra-Laplacian matrices required to study fractional diffusion in any multiplex network. Thus, following Refs.~\cite{riascos14,riascos15,michelitsch19}, we generalize Eq.~(\ref{sol_prob}) as

\begin{equation}
\frac{\mathrm{d}\vec{x}(t) }{\mathrm{d} t}=-\left ( \mathbf{\mathcal{L}}^\mathcal{M} \right )^\gamma \vec{x}(t),
\label{frac_diff}
\end{equation}

\noindent where $\gamma$ is a real number ($0<\gamma<1$) and $\left ( \mathbf{\mathcal{L}}^\mathcal{M} \right )^\gamma$, the combinatorial supra-Laplacian matrix raised to a power $\gamma$, denotes here the \textit{fractional (combinatorial) supra-Laplacian} matrix.

Let us briefly discuss some mathematical properties of the model defined by Eq. (\ref{frac_diff}), as well as qualitative aspects of the expected behavior, limiting cases, and relations to other scenarios characterized by anomalous diffusion. An immediate consequence is that we recover Eq.~(\ref{sol_prob}) in the limit $\gamma\rightarrow 1$. This way of defining the fractional supra-Laplacian matrix preserves the essential features of Laplacian matrices, namely: $\left ( \mathbf{\mathcal{L}}^\mathcal{M} \right )^\gamma$ is (i) positive semidefinite, (ii) stochastic, and (iii) all its non-diagonal elements are non-positive. On the other hand, by setting $D_\alpha=1$ and $D_{\alpha\beta}=1$ (with $\alpha,\beta\in\left \{ 1,\cdots,M \right \}$ and $\beta\neq \alpha$), $\left ( \mathbf{\mathcal{L}}^\mathcal{M} \right )^\gamma$ is equivalent to the fractional Laplacian matrices of monolayer networks described in \cite{riascos14,riascos15,michelitsch19}. For such cases, it has been shown analytically that the continuum limits of the fractional Laplacian matrix (with $0<\gamma<1$) are connected with the operators of fractional calculus. Indeed, in the case of cycle graphs and its continuum limits, the distributional representations for fractional Laplacian matrices take the forms of Riesz fractional derivatives (see Chapter 6 of \cite{michelitsch19} for further details). Besides that, when the above definition of fractional Laplacian matrix is considered, the asymptotic behavior of node-centric CTRWs on homogeneous networks and their continuum limits (with homogeneous and isotropic node distributions) shows explicitly the convergence to a L\'evy propagator associated with the emergence of L\'evy flights with self-similar inverse power-law distributed long-range steps and anomalous diffusion (see Chapter 8 of \cite{michelitsch19} for further details). Alternatively, by using (non-fractional) Laplacian matrices (i.e., $\gamma=1$), Brownian motion (Rayleigh flights) and Gaussian diffusion appear. Both types of asymptotic behaviors are in good agreement with the findings presented in Ref.~\cite{Metzler2000} for the CTRW model with Poisson distribution of waiting times in homogeneus, isotropic systems, when a L\'evy distribution of jump lengths and a Gaussian one are considered, respectively.

We can obtain a set of eigenvalues $\mu_j$ and eigenvectors $\vec{\psi}_j$ of $\mathbf{\mathcal{L}}^\mathcal{M}$ that satisfy the eigenvalue equation $\mathbf{\mathcal{L}}^\mathcal{M} \vec{\psi}_j = \mu_j \vec{\psi}_j$ for $j\in\left \{ 1,\cdots,NM \right \}$ and the orthonormalization condition $\vec{\psi}_i^T \vec{\psi}_j=\delta _{ij}$. Since $\mathbf{\mathcal{L}}^\mathcal{M}$ is a symmetric matrix, the eigenvalues $\mu_j$ are real and nonnegative. In the case of connected multiplex networks, the smallest eigenvalue is 0 and all others are positive.

Following Refs.~\cite{riascos14,riascos15,michelitsch19}, we define the orthonormal matrix $\mathbf{Q}$ with elements $\mathbf{Q}_{ij} = \vec{i}\,^T\vec{\psi}_j $ and the diagonal matrix $\mathbf{\Delta} = \mathrm{diag}\left ( \mu _1,\mu _2,\cdots,\mu _{NM} \right )$. These matrices satisfy $\mathbf{\mathcal{L}}^\mathcal{M}\mathbf{Q}=\mathbf{Q\Delta}$, therefore $\mathbf{\mathcal{L}}^\mathcal{M}=\mathbf{Q\Delta}\mathbf{Q}^\dagger$, where the matrix $\mathbf{Q}^\dagger$ is the conjugate transpose (or Hermitian transpose) of $\mathbf{Q}$. Therefore, we have:

\begin{equation}
\left ( \mathbf{\mathcal{L}}^\mathcal{M} \right )^\gamma = \mathbf{Q\Delta}^\gamma\mathbf{Q}^\dagger=
\sum _{m=1}^{NM} \mu_m^\gamma \vec{\psi}_j\vec{\psi}_j^\dagger ,
\label{frac_Lapl}
\end{equation}

\noindent where $\mathbf{\Delta}^\gamma= \mathrm{diag}\left ( \mu_1^\gamma,\mu_2^\gamma,\cdots,\mu_{NM}^\gamma \right )$. According to Eq.~(\ref{frac_Lapl}), $\left ( \mathbf{\mathcal{L}}^\mathcal{M} \right )^\gamma \vec{\psi}_j = \mu_j^\gamma \vec{\psi}_j$ for $j\in\left \{ 1,\cdots,NM \right \}$. Consequently, the eigenvalues of $\left ( \mathbf{\mathcal{L}}^\mathcal{M} \right )^\gamma$ are equal to those of $\mathbf{\mathcal{L}}^\mathcal{M}$ to the power of gamma, $\mu_j^\gamma$, and the eigenvectors $\vec{\psi}_j$ remain the same for both the supra-Laplacian and the fractional supra-Laplacian matrices.

On the other hand, the diagonal elements of the fractional supra-Laplacian matrix defined in Eq.~(\ref{frac_Lapl}) introduce a generalization of the strength $\sigma_i^\alpha = \mathbf{\mathcal{L}}^\mathcal{M}_{ff}$ with $f=i+(\alpha-1)N$ to the fractional case. In this way, the fractional strength of node $i$ at layer $\alpha$ is given by:

\begin{equation}
\left ( \sigma_i^\alpha \right )^{(\gamma)}=\left ( \mathbf{\mathcal{L}}^\mathcal{M} \right )_{ff}^\gamma =
\sum _{m=1}^{NM} \mu_m^\gamma \mathbf{Q}_{fm}\mathbf{Q}_{fm}^*,
\label{frac_degree}
\end{equation}

\noindent where $X^*$ denotes the conjugate of $X$. In general, the elements of the fractional (combinatorial) supra-Laplacian matrix can be calculated as follows:

\begin{equation}
\left ( \mathbf{\mathcal{L}}^\mathcal{M} \right )_{fg}^\gamma =\sum _{m=1}^{2N} \mu_m^\gamma \mathbf{Q}_{fm}\mathbf{Q}_{gm}^*.
\label{element_frac_Lapl}
\end{equation}

Now, by analogy with the random walk normalized supra-Laplacian matrix $\mathcalboondox{L}=\mathbf{\mathcal{S}}^{-1}\mathbf{\mathcal{L}}^\mathcal{M}$, we introduce the normalized
fractional supra-Laplacian matrix $\mathcalboondox{L}^{\left (\gamma \right )}$ with elements

\begin{equation}
\mathcalboondox{L}_{fg}^{\left (\gamma \right )}=\frac{\left ( \mathbf{\mathcal{L}}^\mathcal{M} \right )_{fg}^\gamma}{\left ( \mathbf{\mathcal{L}}^\mathcal{M} \right )_{ff}^\gamma}=\delta_{fg}-\mathcalboondox{T}_{fg}^{\left (\gamma \right )},
\label{frac_RWLapl}
\end{equation}

\noindent where $\mathcalboondox{T}_{fg}^{\left (\gamma \right )}$ denotes the elements of the $NM\times NM$ fractional transition matrix $\mathcalboondox{T}^{\left (\gamma \right )}$. Note that $\mathcalboondox{T}^{\left (\gamma \right )}$ is a stochastic matrix, that satisfies $\mathcalboondox{T}_{ff}^{\left (\gamma \right )}=0$ and $\sum_{g=1}^{NM} \mathcalboondox{T}_{fg}^{\left (\gamma \right )}=1$.

Finally, when fractional diffusion takes place for a given $\gamma$, the probability $p(t)_{i^\alpha\rightarrow j^\beta}^{\left (\gamma \right )}$ of finding a node-centric CTRW at node $j$ in the $\beta$th layer (at time $t$), when the random walker was initially located at node $i$ in the $\alpha$th layer, is expressed by:

\begin{equation}
p(t)_{i^\alpha\rightarrow j^\beta}^{\left (\gamma \right )}=\vec{p}(t)_g^{\left (\gamma \right )}= \vec{\mathcal{C}}_g\,^T \exp\left (-t\mathcalboondox{L}^{(\gamma)} \right ) \vec{\mathcal{C}}_f,
\label{prob_walker_frac}
\end{equation}

\noindent where $g=j+(\beta-1)N$ and $f=i+(\alpha-1)N$ with $i,j \in \left \{ 1, \cdots, N \right \}$ and $\alpha,\beta \in \left \{ 1, \cdots, M \right \}$. Thus, the MSD for fractional dynamics, denoted as $\left \langle r^2(t) \right \rangle^{\left (\gamma \right )}$, is given by:

\begin{equation}
\left \langle r^2(t) \right \rangle^{\left (\gamma \right )} = \frac{1}{N^2M^2}\sum_{\alpha=1}^M \sum_{\beta=1}^{M} \sum_{i=1}^N \sum_{j=1}^{N}\left ( d_{i^\alpha\rightarrow j^\beta} \right )^2 p(t)_{i^\alpha\rightarrow j^\beta}^{\left (\gamma \right )}.
\label{eq_frac_MSD_multpl}
\end{equation}

\noindent According to Eq.~(\ref{eq_frac_MSD_multpl}), the time evolution of $\left \langle r^2(t) \right \rangle^{\left (\gamma \right )}$ and the corresponding diffusive behavior for the multiplexes $\mathcal{M}$ considered in this work depend on $\gamma$, the number of nodes $N$, the total amount of layers $M$, the topology of each layer, given by $\mathbf{A}^\alpha$, and the intra- and inter-layer diffusion constants $D_\alpha$ and $D_{\alpha\beta}$ (with $\alpha,\beta\in\left \{ 1,\cdots,M \right \}$ and $\beta\neq \alpha$), respectively. As expected, when $\gamma\rightarrow 1$, Eqs.~(\ref{frac_degree}), (\ref{frac_RWLapl}), (\ref{prob_walker_frac}), and  (\ref{eq_frac_MSD_multpl}) reproduce the corresponding equations in the previous section.

\subsection{Circulant multiplexes}
\label{Sec:Circulant_Analitycal}

In this subsection we analyze fractional diffusion on a $N$-node multiplex network in which all its $M$ layers consist of  \textit{interacting cycle graphs} i.e., each layer $\alpha$ (with $\alpha \in \left \{ 1, \cdots, M \right \}$) contains a $N-$ring topology in which each node is connected to its $J^\alpha$ left and $J^\alpha$ right nearest nodes. Thus, $J^\alpha$ represent the interaction parameter of the layer $\alpha$. It is easy to see that, if $N$ is odd, then $J^\alpha \in \left \{ 1, \cdots, (N-1)/2 \right \}$. Note that, when $J^\alpha=1$, the $\alpha$-th  layer contains a cycle graph whereas, if  $J^\alpha=(N-1)/2$, it corresponds to a complete graph. For the purpose of deriving exact expressions for the eigenvalues, hearafter we only consider multiplex networks with $M=2$ layers and odd number of nodes. Besides, to emphasize the inter-layer diffusion process and simplify the notation, we choose the diffusion coefficients $D_1=D_2=1$ and $D_{12}/D_\alpha=D_x$ for $\alpha \in \left \{ 1, 2 \right \}$ \cite{gomez13,cencetti19,tejedor18}.

According to Eqs.~(\ref{def_comb_lap})-(\ref{interlayer_connect_matrix}), the combinatorial supra-Laplacian matrix of the multiplex is written as

\begin{equation}
\mathbf{\mathcal{L}}^\mathcal{M}=\left ( \begin{matrix}\mathbf{L}^1 +D_x\mathbf{I}_N & -D_x\mathbf{I}_N\\
-D_x\mathbf{I}_N & \mathbf{L}^2 +D_x\mathbf{I}_N
\end{matrix} \right )
=\left ( \begin{matrix}\mathbf{C}^1 & -D_x\mathbf{I}_N\\
-D_x\mathbf{I}_N & \mathbf{C}^2
\end{matrix} \right ),
\label{comb_circ_ini1}
\end{equation}

\noindent where both $\mathbf{C}^1$ and $\mathbf{C}^2$ are $N\times N$ circulant matrices. Since exact analytical expressions for the eigenvalues and eigenvectors of circulant matrices are well known \cite{Mieghem11}, it is also possible to obtain similar expressions for $\mu_j$ and $\vec{\psi}_j$ (for $j\in\left \{ 1,\cdots,2N \right \}$). So let us write

\begin{equation}
\mathbf{F}^{-1}
\left ( \begin{matrix}\mathbf{C}^1 & -D_x\mathbf{I}_N\\
-D_x\mathbf{I}_N & \mathbf{C}^2
\end{matrix} \right ) \mathbf{F}
=
\left ( \begin{matrix}\mathrm{\Xi}^1 & -D_x\mathbf{I}_N\\
-D_x\mathbf{I}_N & \mathrm{\Xi}^2
\end{matrix} \right )
\label{comb_circ_ini2}
\end{equation}

\noindent where

\begin{equation}
\mathbf{F} = \left ( \begin{matrix} \mathbf{U} & \mathbf{0}\\
\mathbf{0} & \mathbf{U}
\end{matrix} \right )
\label{def_F}
\end{equation}

\noindent is a $2N\times 2N$ block-diagonal matrix, $\mathbf{U}$ is the $N\times N$ hermitian matrix with elements $\mathbf{U}_{ij}=\omega ^{(i-1)(j-1)}/\sqrt{N}$, $\omega \equiv \exp(-\mathfrak{i}2\pi/N)$, $\mathfrak{i}=\sqrt{-1}$, $\mathrm{\Xi}^\alpha=\mathrm{diag}\left (  \xi_1^\alpha ,\cdots, \xi_{N}^\alpha  \right )$, and $\xi_m^\alpha$ are the eigenvalues of $\mathbf{C}^\alpha$, given by

\begin{align}
\xi_m^\alpha&=D_x+A_m^\alpha\equiv D_x+2\left (J^\alpha+1\right )-\frac{2\sin\left ((J^\alpha +1)\frac{\pi(m-1)}{N}\right )\cos\left (J^\alpha\frac{\pi(m-1)}{N}\right )}{\sin\left (\frac{\pi(m-1)}{N}\right )}
\label{eigen_xi}
\end{align}

\noindent for $1<m\leq N$, and $\xi_m^\alpha=D_x$ for $m=1$. Since the matrices $-D_x\mathbf{I}_N$ and $\left (\mathrm{\Xi}^2-\mu_m\mathbf{I}_N\right )$ commute, the eigenvalues of $\mathbf{\mathcal{L}}^\mathcal{M}$ can be obtained as:

\begin{equation}
\mu_{2m-1}= \frac{\xi_m^1+\xi_m^2+\sqrt{\left (\xi_m^1-\xi_m^2\right )^2+4D_x^2}}{2},
\label{lapl_eigenA}
\end{equation}

\noindent and

\begin{equation}
\mu_{2m}=\frac{\xi_m^1+\xi_m^2-\sqrt{\left (\xi_m^1-\xi_m^2\right )^2+4D_x^2}}{2},
\label{lapl_eigenB}
\end{equation}

\noindent for $m\in\left \{ 1,\cdots,N \right \}$. Note that the eigenvalues $\mu_m$ are not ordered from smallest to largest and vice versa (for instance, when $m=1$, $\mu_2=0$). Given such set of eigenvalues, the corresponding hermitian matrix of eigenvectors $\mathbf{Q}=\left ( \begin{matrix} \vec{\psi}_1 &\cdots  & \vec{\psi}_{2N}\end{matrix} \right )$ has the following elements:

\begin{widetext}

\begin{equation}
\left.\begin{matrix}
\mathbf{Q}_{fg}=\frac{1}{\sqrt{N\left ( 1+M_{g}^2 \right )}}\exp\left ( -\mathfrak{i}\frac{2\pi}{N} (f-1)\left \lfloor \frac{g-1}{2} \right \rfloor \right)\\
\mathbf{Q}_{(f+N)g}=\frac{M_g}{\sqrt{N\left ( 1+M_{g}^2 \right )}}\exp\left ( -\mathfrak{i}\frac{2\pi}{N} (f-1)\left (\left \lfloor \frac{g-1}{2} \right \rfloor +N \right ) \right)
\end{matrix}\right\}\mathrm{for}\;f,g\in\left \{ 1,\cdots,N \right \},
\label{Q_layer1}
\end{equation}

\noindent and

\begin{equation}
\left.\begin{matrix}
\mathbf{Q}_{(f-N)g}=\frac{1}{\sqrt{N\left ( 1+M_{g}^2 \right )}}\exp\left ( -\mathfrak{i}\frac{2\pi}{N}(f-1-N)\left (\left \lfloor \frac{g-1}{2} \right \rfloor +N \right ) \right)\\
\mathbf{Q}_{fg}=\frac{M_g}{\sqrt{N\left ( 1+M_{g}^2 \right )}}\exp\left ( -\mathfrak{i}\frac{2\pi}{N}(f-1)\left (\left \lfloor \frac{g-1}{2} \right \rfloor +N\right ) \right)
\end{matrix}\right\}\mathrm{for}\;f,g \in \left \{ N+1,\cdots,2N \right \}
\label{Q_layer2}
\end{equation}

\end{widetext}

\noindent where

\begin{equation}
M_g=\frac{\xi_{\left ( 1+\left \lfloor (g-1)/2 \right \rfloor\right)}^1-\mu_g}{D_x},
\label{coeff_M_g}
\end{equation}

\noindent and $\left \lfloor . \right \rfloor$ denotes the floor function [see Appendix A for further details on the derivation of Eqs.~(\ref{lapl_eigenA})-(\ref{coeff_M_g})].

Using the eigenvalue spectrum of $\mathbf{\mathcal{L}}^\mathcal{M}$ [Eqs.~(\ref{lapl_eigenA}) and (\ref{lapl_eigenB})] and its eigenvectors [Eqs.~(\ref{Q_layer1}) and (\ref{Q_layer2})], the fractional strength of any node at layer 1 is

\begin{align}
\sigma_1^{(\gamma)}&\equiv\left ( \sigma_i^1 \right )^{(\gamma)}=\left ( \mathbf{\mathcal{L}}^\mathcal{M} \right )_{ii}^\gamma =\sum _{m=1}^{2N} \mu_m^\gamma \mathbf{Q}_{im}\mathbf{Q}_{im}^* =\sum _{m=1}^{2N} \frac{1}{N\left ( 1+M_{m}^2 \right )}\mu_{m}^\gamma,
\label{frac_degree1}
\end{align}

\noindent whereas the fractional strength of the nodes at layer 2 is given by

\begin{align}
\sigma_2^{(\gamma)}&\equiv\left ( \sigma_i^2 \right )^{(\gamma)}=\left ( \mathbf{\mathcal{L}}^\mathcal{M} \right )_{(i+N)(i+N)}^\gamma=\sum _{m=1}^{2N} \mu_m^\gamma \mathbf{Q}_{(i+N)m}\mathbf{Q}_{(i+N)m}^*
=\sum _{m=1}^{2N} \frac{M_{m}^2}{N\left ( 1+M_{m}^2 \right )}\mu_{m}^\gamma,
\label{frac_degree2}
\end{align}

\noindent for $i \in \left \{ 1,\cdots,N \right \}$. Note that Eqs.~(\ref{frac_degree1}) and (\ref{frac_degree2}) do not depend on $i$, as expected for circulant layers of interacting cycles.

The set of Eqs.~(\ref{element_frac_Lapl})-(\ref{eq_frac_MSD_multpl}) and (\ref{eigen_xi})-(\ref{frac_degree2}) allows to derive the the fractional (combinatorial) supra-Laplacian matrix ($\left ( \mathbf{\mathcal{L}}^\mathcal{M} \right )^\gamma$), the normalized fractional supra-Laplacian matrix ($\mathcalboondox{L}^{\left (\gamma \right )}$), the fractional transition matrix ($\mathcalboondox{T}^{\left (\gamma \right )}$), the probability of finding a node-centric CTRW at a given position ($p(t)_{i^\alpha\rightarrow j^\beta}^{\left (\gamma \right )}$), and $\left \langle r^2(t) \right \rangle^{\left (\gamma \right )}$.

Finally, note that, according to Eq.~(\ref{frac_Lapl}), the fractional supra-Laplacian matrix $\left ( \mathbf{\mathcal{L}}^\mathcal{M} \right )^\gamma$ is a block-matrix whose blocks are circulant and, consequently, so is the normalized fractional supra-Laplacian matrix $\mathcalboondox{L}^{\left (\gamma \right )}$. Therefore, using a strategy similar to that previously described for $\mathbf{\mathcal{L}}^\mathcal{M}$, it is possible to derive the eigenvalue spectrum of $\mathcalboondox{L}^{\left (\gamma \right )}$. After conducting the necessary manipulation, the resulting eigenvalues are given by:

\begin{equation}
\lambda_{2m-1}= \frac{\kappa+\sqrt{\kappa^2-\nu}}{2},
\label{RWlapl_eigenA}
\end{equation}

\noindent and

\begin{equation}
\lambda_{2m}=\frac{\kappa-\sqrt{\kappa^2-\nu}}{2},
\label{RWlapl_eigenB}
\end{equation}

\noindent where

\begin{align}
\kappa &=\frac{\mu_{2m-1}^\gamma  }{\left ( 1+\left ( S + C \right )^2 \right )}\left ( \frac{1}{\sigma_1^{(\gamma )}}+ \frac{\left ( S + C \right )^2}{\sigma_2^{(\gamma )}}\right )+\frac{\mu_{2m}^\gamma}{\left ( 1+\left ( S - C \right )^2 \right )}\left ( \frac{1}{\sigma_1^{(\gamma )}}+ \frac{\left ( S - C \right )^2}{\sigma_2^{(\gamma )}}\right ),
\end{align}

\begin{align}
\nu=\frac{16C^2}{\sigma_1^{(\gamma )}\sigma_2^{(\gamma )}}\frac{\mu_{2m-1}^\gamma  \mu_{2m}^\gamma}{\left ( 1+\left ( S + C \right )^2 \right )\left ( 1+\left ( S - C \right )^2 \right )},
\label{nu_def}
\end{align}

\begin{align}
S=\frac{A_m^1-A_m^2}{2D_x},
\label{def_S}
\end{align}

\noindent and $C^2=1+S^2$ for $m \in \left \{ 1,\cdots,N \right \}$ [see Appendix B for further details on the derivation of Eqs.~(\ref{RWlapl_eigenA})-(\ref{def_S})]. Note that the eigenvalues $\mu_m$ and $\mu_m^\gamma$ of $\mathbf{\mathcal{L}}^\mathcal{M}$ and $\left ( \mathbf{\mathcal{L}}^\mathcal{M} \right )^\gamma$ are not ordered and, consequently, neither are the $\lambda_m$. It is also noteworthy that  $\lambda_m$ (with $m \in \left \{ 1,\cdots,2N \right \}$) depend on the value of the inter-layer diffusion constant, $D_x$.

By using Eqs.~(\ref{RWlapl_eigenA}) and (\ref{RWlapl_eigenB}), it is possible to obtain the algebraic connectivity of $\mathcalboondox{L}^{\left (\gamma \right )}$, i.e., its second-smallest eigenvalue, denoted here as $\Lambda_2$.  When $D_x\rightarrow 0$, $\Lambda_2=\lambda_1$, that is:

\begin{equation}
\Lambda_2=\frac{1}{2}\left ( 2D_x \right )^\gamma\left ( \frac{1}{\sigma_1^{(\gamma )}}+\frac{1}{\sigma_2^{(\gamma )}} \right ) \sim
D_x ^{\gamma}.
\label{alge_conn_small}
\end{equation}

\noindent In the case of $\Lambda_2 \neq \lambda_1$, it is easy to see that  $\Lambda_2 = \lambda_{2m} < \lambda_{2m-1}$ for $m \in \left \{ 1,\cdots,N \right \}$ [see Eqs.~(\ref{RWlapl_eigenA}) and (\ref{RWlapl_eigenB})]. Thus, for $J^1\neq J^2$ and $D_x\rightarrow \infty$, the algebraic connectivity can be approximated as:

\begin{equation}
\Lambda_2\approx\left ( \frac{A_{2m}^1+A_{2m}^2}{2D_x} \right )^\gamma \sim D_x ^{-\gamma}.
\label{alge_conn_large}
\end{equation}

\noindent According to Eqs.~(\ref{alge_conn_small}) and (\ref{alge_conn_large}), the algebraic connectivity of the normalized supra-Laplacian matrix $\mathcalboondox{L}^{\left (\gamma \right )}$ is nonmonotonic. When $D_x\rightarrow0$, the inter-layer diffusion disappears and the dynamics reduces to those for diffusion on single (isolated) layers. On the other hand, when $D_x\rightarrow\infty$, the strength of the vertices are approximately equal to interlayer connection, i.e., $\sigma_i^{(\gamma)}\approx D_x^\gamma$. Thus, the centric-node random walkers spend most of the time in switching layer, instead of jumping to other vertices. Consequently, when the interlayer diffusion coefficient is very small or very large, the diffusion is hindered. For that reason, there is an optimal range of values for $D_x$. Note that the previous nonmonotonic trends of node-centric CTRWs emerge from the multiplex structure itself and, consequently, they persist even in the case of $\gamma=1$.


\section{Results}
\label{Sec:results}

In this section, we present our results for the diffusion processes of Poissonian node-centric CTRWs on multiplex networks. Our discussion is mainly focused on their rate of convergence to the steady state in such systems, given by the diffusion time scale $\tau\sim1/\Lambda_2$ \cite{masuda17}.  To do so, we analyze the nonmonotonic dependence of $\Lambda_2$ on the inter-layer coupling (i.e., $D_x$), as well as the influence of fractional dynamics (i.e., $\gamma$). In subsection \ref{regular_multiplex}, we present our analytical results for $\Lambda_2$ in regular multiplex networks. By analyzing the case of regular multiplexes with $J=1$, in subsection \ref{long_navigation} we show that the enhanced diffusion induced by fractional dynamics is due to the emergence of a new type long-range navigation between layers. In subsection \ref{sec_MSD}, we provide an example of optimal convergence to the steady state of node-centric CTRWs, discussing the dependence of $\left \langle r^2(t) \right \rangle^{(\gamma)}$ on $D_x$ and $\gamma$. Finally, in subsection \ref{sec_nonreg} we present our results for $\Lambda_2$ in non-regular multiplexes.


\subsection{Regular multiplexes}
\label{regular_multiplex}

Regular multiplex networks meet the condition $J\equiv J^1=J^2$. Therefore, in these systems $\xi_m\equiv \xi_m^1=\xi_m^2=D_x+A_m$ with $A_m\equiv A_m^1=A_m^2$ and $m\in\left \{ 1,\cdots,N \right \}$.  Under these circumstances, the eigenvalues of $\mathbf{\mathcal{L}}^\mathcal{M}$ reduce to $\mu_{2m-1}= A_m+2D_x$ and $\mu_{2m}= A_m$ (for $m\in\left \{ 1,\cdots,N \right \}$). Thus, according to Eq.~(\ref{coeff_M_g}), $(M_m)^2=1$ and the fractional strength of the nodes of both layers [see Eqs.~(\ref{frac_degree1}) and (\ref{frac_degree2})] is given by:

\begin{align}
\sigma^{(\gamma)}&\equiv \sigma_1^{(\gamma)}=  \sigma_2^{(\gamma)} =\sum _{m=1}^{N} \frac{1}{2N}\left ( \mu_{2m-1}^\gamma +\mu_{2m}^\gamma\right )=\frac{1}{2N}\sum _{m=1}^{N} \left ( \left ( A_m+2D_x\right )^\gamma + \left ( A_m\right )^\gamma\right ).
\label{frac_degree_reg}
\end{align}

\begin{figure}[h!]
\centering
\subfloat[]{
\centering
\includegraphics[width=0.5\linewidth]{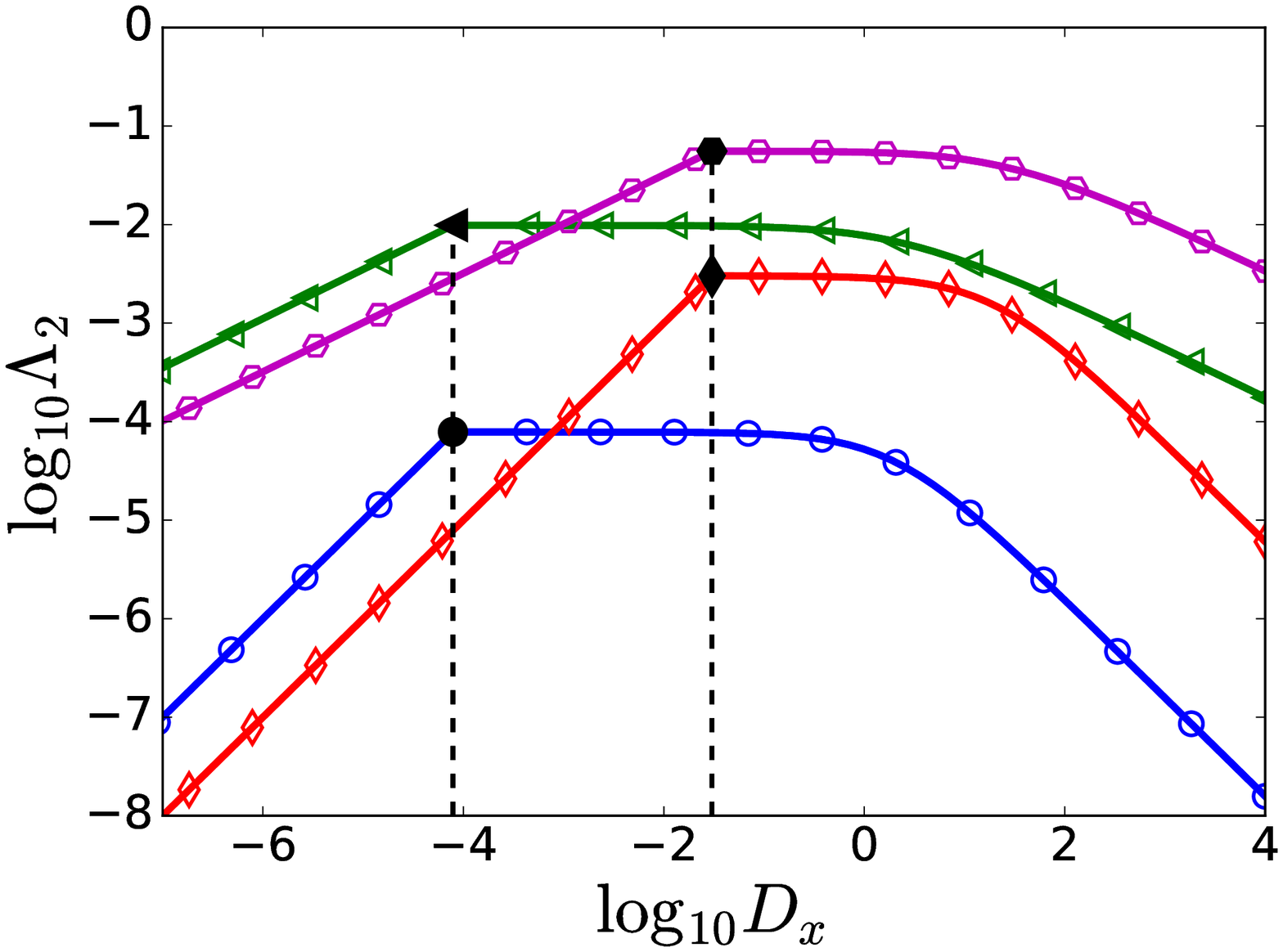}
\label{}
}
\subfloat[]{
\centering
\includegraphics[width=0.5\linewidth]{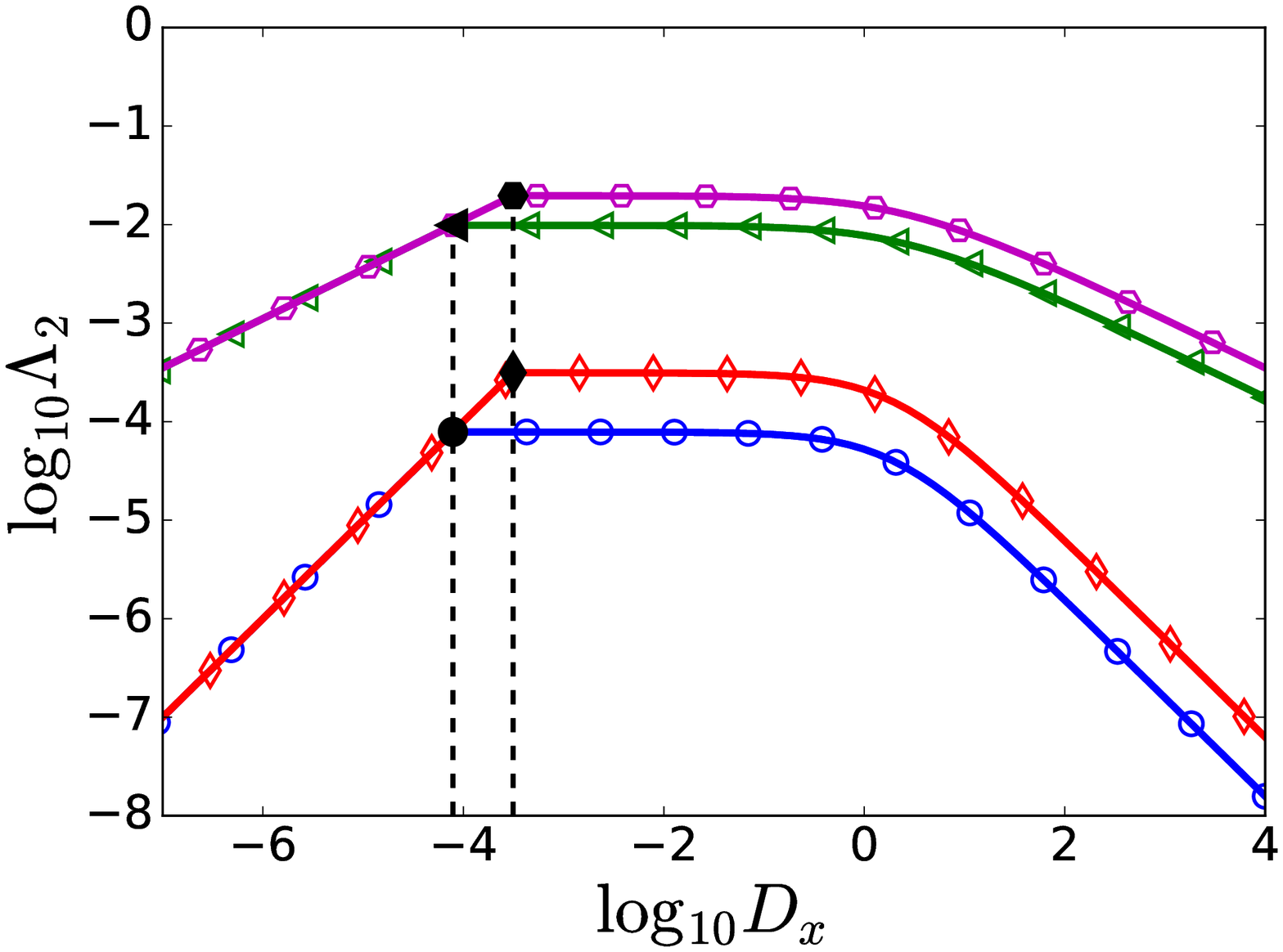}
\label{}
}
\caption{Dependence of $\Lambda_2$ on $D_x$ for regular multiplex networks with two layers. (a) Results for $N=501$ nodes: $J=1$ and $\gamma=1$ (blue circles), $J=10$ and $\gamma=1$ (red diamonds), $J=1$ and $\gamma=0.5$ (green triangles), $J=10$ and $\gamma=0.5$ (magenta hexagons). (b) Results for $J=1$: $N=501$ and $\gamma=1$ (blue circles), $N=251$ and $\gamma=1$ (red diamonds), $N=501$ and $\gamma=0.5$ (green triangles), and $N=251$ and $\gamma=0.5$ (magenta hexagons). Hollow symbols represent the results obtained from computer simulations, while continuous lines show the findings obtained from Eq.~(\ref{algebraic_reg}). Solid black symbols indicate the optimal value $D_x = {A_c}/{2}$ highlighting that, for fixed $J$ and $N$, they do not depend on $\gamma$.}
\label{lambda_2_reg}
\end{figure}

On the other hand, since the fractional strength is a constant for all nodes, the eigenvalues of the normalized supra-Laplacian matrix are given by $\lambda_m=\mu_m/\sigma^{(\gamma)}$ for $m\in\left \{ 1,\cdots,2N \right \}$, while the matrix of the corresponding eigenvectors is $\mathbf{Q}$ [see Eqs.~(\ref{Q_layer1}) and (\ref{Q_layer2})]. Thus, in regular multiplexes the algebraic connectivity can be calculated as:

\begin{equation}
\Lambda_2=
\left\{
\begin{matrix}
\left ( 2D_x \right )^\gamma /\sigma^{(\gamma )} &\mathrm{for}\; D_x\leq {A_c}/{2}, \\
A_c^\gamma /\sigma^{(\gamma)}  & \mathrm{for}\; D_x \geq  {A_c}/{2},
\end{matrix}
\right.
\label{algebraic_reg}
\end{equation}

\noindent where $c\in \left \{ 2,N \right \}$ represents the natural number that minimizes $A_m$ [see Eq.~(\ref{eigen_xi})]. According to Eq.~(\ref{algebraic_reg}), $\Lambda_2$ reaches a global maximum at $D_x = {A_c}/{2}$. Therefore, such value of the interlayer diffusion constant guarantees the fastest convergence to the steady state of node-centric CTRWs in these regular multiplexes (for an example see subsection \ref{sec_MSD}). On the other hand, it is worth mentioning that $D_x = {A_c}/{2}$ does not depend on $\gamma$. Therefore, for a given $J$ and $N$, the optimal value remains the same. However, the smaller the $\gamma$, the larger the optimal algebraic connectivity. Thus, for a given value of $N$ and $J$, the more intense the fractional diffusion (i.e., the smaller the paramenter $\gamma$), the larger $\Lambda_2$ is and, therefore, the faster the convergence to the steady state is ($\tau\sim1/\Lambda_2$). In Fig.~\ref{lambda_2_reg}(a), we show the dependence of $\Lambda_2$ on $D_x$ for several regular multiplex networks with two layers. As can be observed, the numerical results are in excelent agreement with  Eq.~(\ref{algebraic_reg}). Finally, note that, according to Eqs.~(\ref{eigen_xi}) and (\ref{algebraic_reg}), for $c=2$ and fixed values of $J$ and $\gamma$, the smaller the system size $N$, the larger $D_x=A_c/2$ and $\Lambda_2$, corroborating the expected results of a faster convergence to the steady state.

\subsection{Emergence of interlayer long-range navigation}
\label{long_navigation}

In this section we explore the navigation strategy of node-centric CTRWs on regular multiplex networks that are formed by cycle graphs (i.e., $J=1$). We will explore the probability transition between two nodes $i$ and $j$ that are located in different layers. Let us suppose that $i$ is at layer 1 and $j$ is at layer 2. Following Refs. \cite{riascos14,riascos15,michelitsch19}, by using Eqs.~(\ref{Q_layer1})-(\ref{coeff_M_g}), it is possible to approximate the element of the fractional supra-Laplacian that refers to $i$ and $j$ as:

\begin{align}
\left ( \mathbf{\mathcal{L}}^\mathcal{M} \right )_{i(j+N)}^\gamma
\approx \frac{1}{2N} \sum _{m=1}^{N} \left ( A_m^\gamma -\gamma A_m \left (2D_x \right )^{\gamma-1} \right )\exp\left ( \mathfrak{i}\theta_{m}d\right )-\frac{1}{2N}\sum _{m=1}^{N} \left ( 2D_x \right )^\gamma \exp\left ( \mathfrak{i}\theta_{m}d\right ),
\label{element_frac_Lapl_regular}
\end{align}

\noindent for $\left | D_x \right |>\left | A_m \right |$ and $D_x>0$, where $d\equiv d_{i^1\rightarrow j^1}$, $\theta_m=2\pi\left ( m-1 \right )/N$ represents the shortest path distance between $i$ and $j$ at layer 1, and $A_m=2+2\cos\left ( 2\theta_m\right )$ [see Eq.(\ref{eigen_xi}) for $J^1=J^2=1$ and Appendix C for further details of these derivations]. Besides that, note that $\left ( \mathbf{\mathcal{L}}^\mathcal{M} \right )_{i(j+N)}^\gamma=0$ for $D_x=0$.

On the other hand, by using Eq.~(\ref{frac_degree_reg}), a similar expression can be derived for the fractional strength:

\begin{align}
\sigma^{(\gamma)}
\approx \frac{1}{2N} \sum _{m=1}^{N} \left ( A_m^\gamma +\gamma A_m \left (2D_x \right )^{\gamma-1} \right )\exp\left ( \mathfrak{i}\theta_{m}0\right )+\frac{1}{2N}\sum _{m=1}^{N} \left ( 2D_x \right )^\gamma \exp\left ( \mathfrak{i}\theta_{m}0\right ).
\label{frac_degree_regJ1}
\end{align}

Following Refs. \cite{riascos14,riascos15,michelitsch19}, Eqs.~(\ref{element_frac_Lapl_regular}) and (\ref{frac_degree_regJ1}) can be expressed in terms of an integral in the thermodynamic limit (i.e., $N\rightarrow\infty$) which can be explored analytically (see
Ref. [43] for a discussion on that integral and Appendix C). The resulting expressions are given by:

\begin{align}
\left ( \mathbf{\mathcal{L}}^\mathcal{M} \right )_{i(j+N)}^\gamma
\approx -\frac{1}{2}\gamma \left ( 2D_x \right )^{\gamma-1} K_d-\frac{1}{2}\left ( 2D_x \right )^{\gamma}\delta_{d0}-\frac{1}{2}\frac{\Gamma \left ( d-\gamma  \right )\Gamma \left ( 2\gamma +1 \right )}{\pi \Gamma \left ( 1+\gamma +d \right )}\sin\left ( \pi \gamma  \right ),
\label{element_frac_Lapl_regular_inf}
\end{align}

\noindent and

\begin{align}
\sigma^{(\gamma)}
\approx \gamma \left ( 2D_x  \right )^{\gamma-1} -\frac{1}{2}\frac{\Gamma \left ( -\gamma  \right )\Gamma \left ( 2\gamma +1 \right )}{\pi \Gamma \left ( 1+\gamma \right )}\sin\left ( \pi \gamma  \right )+\frac{1}{2}\left ( 2D_x \right )^{\gamma},
\label{frac_degree_regJ1_inf}
\end{align}

\noindent where

\begin{align}
K_d=\left\{\begin{matrix}
2 & \mathrm{if}\;d=0, \\
-1 & \mathrm{if}\;d=1,\\
0 & \mathrm{otherwise},
\end{matrix}\right.
\label{Eq_Kd}
\end{align}

\noindent and $\Gamma (x)$ is the $\Gamma$ function [see Appendix C for further details on the derivation of Eqs.~(\ref{element_frac_Lapl_regular_inf})-(\ref{Eq_Kd})].

\begin{figure}[h!]
\centering
\subfloat[]{
\centering
\includegraphics[width=0.5\linewidth]{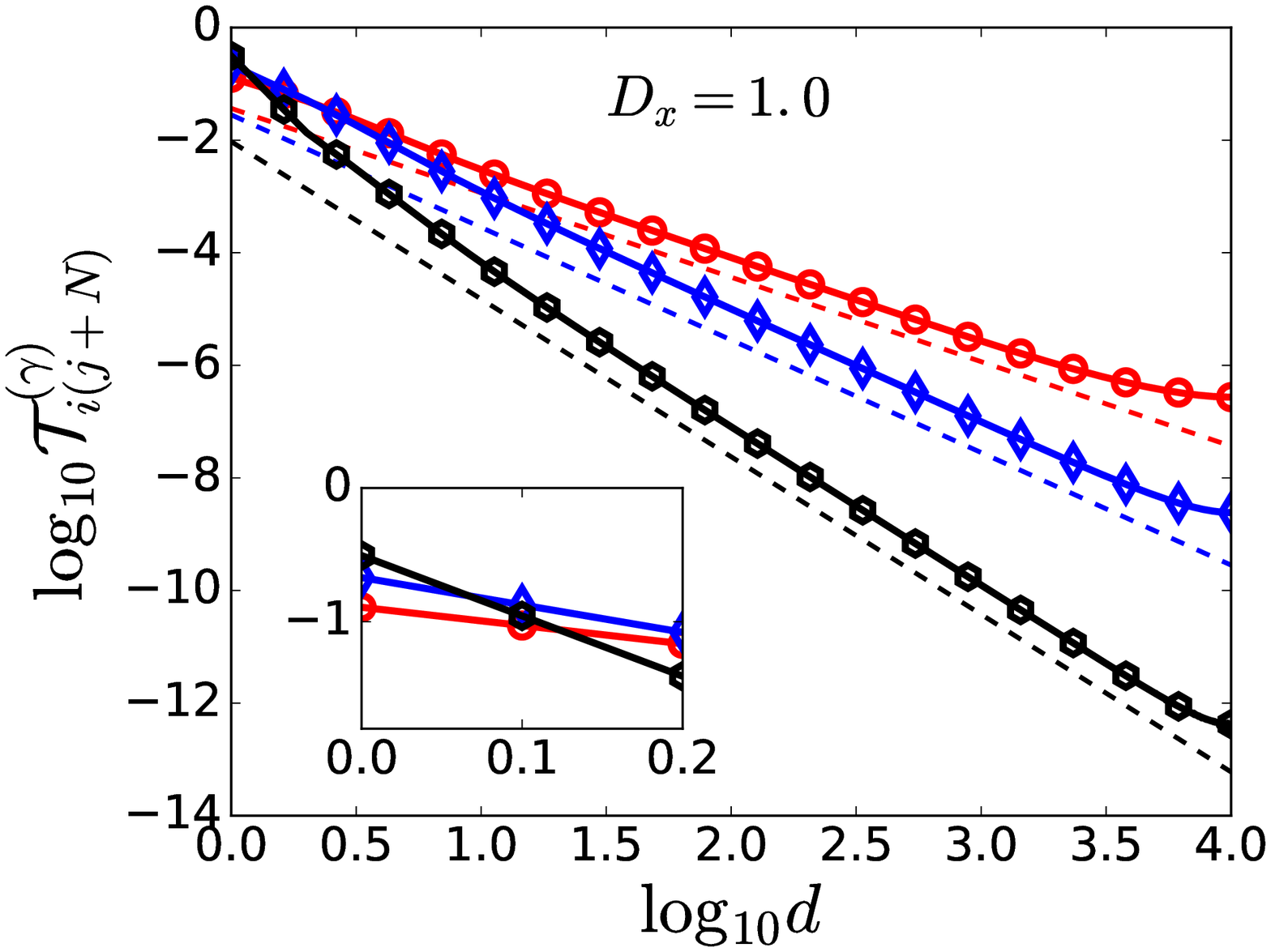}
\label{der_sim_MSD_EqCG}
}
\subfloat[]{
\centering
\includegraphics[width=0.5\linewidth]{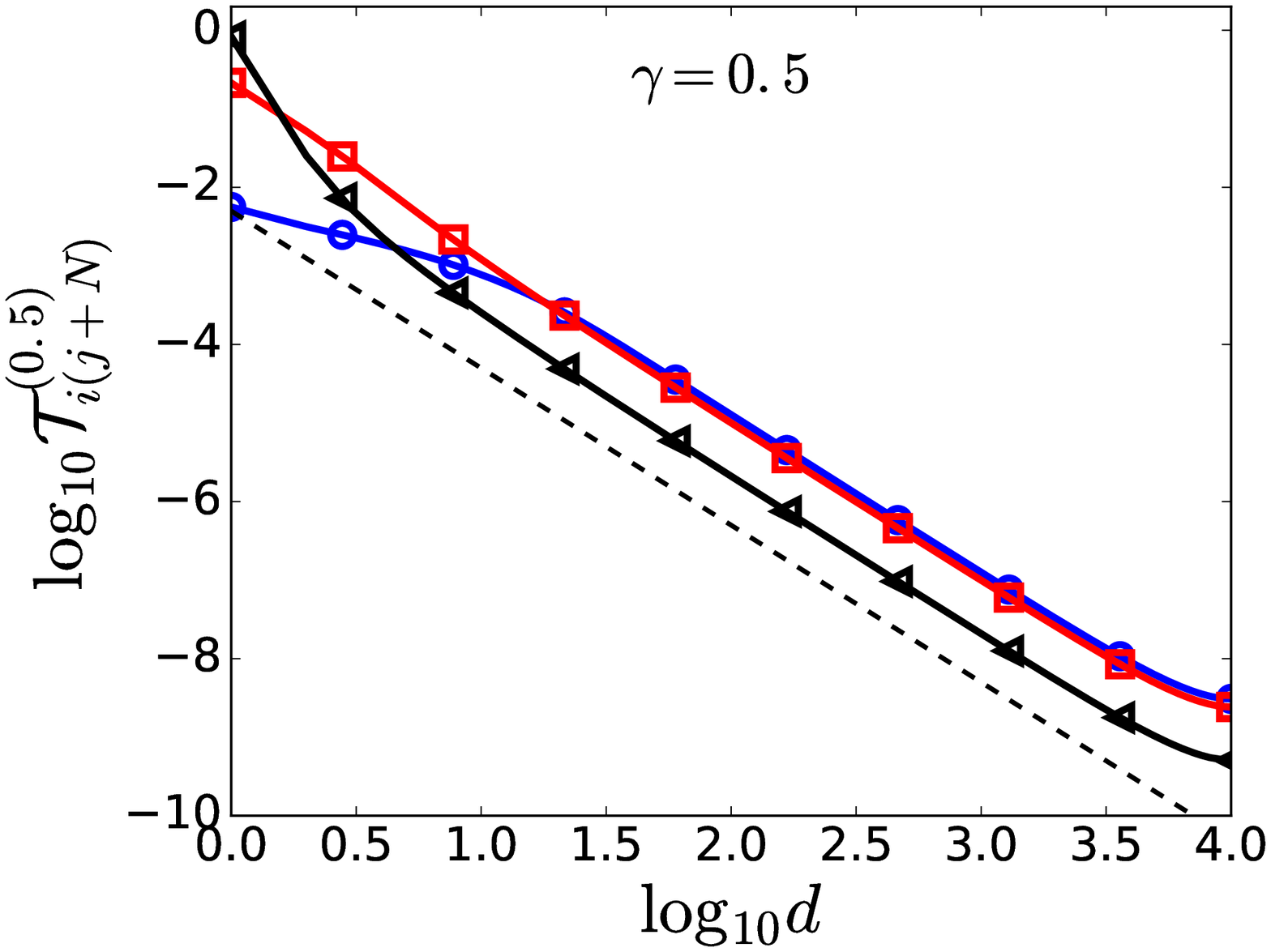}
\label{der_sim_MSD_EqCG}
}
\caption{(a) Dependence of $\mathcalboondox{T}_{i(j+N)}^{(\gamma)}$ on $d$ for regular multiplex networks with two layers, $N=20001$ nodes, $J=1$, and $D_x=2$: $\gamma=0.25$ (red circles), $\gamma=0.5$ (blue diamonds), and $\gamma=0.9$ (black hexagons) [see Eqs.~(\ref{frac_RWLapl})]. Inset: Detail of the previous series when $d\sim1$. (b) Dependence of $\mathcalboondox{T}_{i(j+N)}^{(\gamma)}$ on $D_x$ when $\gamma=0.5$, $N=20001$, and $J=1$: $D_x= 0.01$ (blue circles), $D_x=1$ (red squares), and $D_x= 100$ (black triangles). In both panels, the color dashed lines show power-law decay with exponent $\varrho= 1+2\gamma$ [see Eq (\ref{power_decay})].}
\label{ltransicion_gamma_0p5}
\end{figure}

According to  Eqs.~(\ref{frac_RWLapl}), in the therodynamic limit, the elements of the transition matrix between two nodes $i$ and $j$ that belong to different layers, can be approximated by:

\begin{align}
\mathcalboondox{T}_{i(j+N)}^{(\gamma)}=\delta_{i(j+N)}-\frac{\left ( \mathbf{\mathcal{L}}^\mathcal{M} \right )_{i(j+N)}^\gamma}{\sigma^{(\gamma)}}\approx \frac{1}{2\sigma^{(\gamma)}}\frac{\Gamma \left ( d-\gamma  \right )\Gamma \left ( 2\gamma +1 \right )}{\pi \Gamma \left ( 1+\gamma +d \right )}\sin\left ( \pi \gamma  \right ),
\end{align}

\noindent for $d\gg1$ (i.e., $K_d=0$). By using the asymptotic property $\Gamma \left ( x+\gamma  \right )\approx\Gamma \left ( x \right )x^\gamma$, it is possible to express $\mathcalboondox{T}_{i(j+N)}^{(\gamma)}$ as follows:

\begin{equation}
\mathcalboondox{T}_{i(j+N)}^{(\gamma)} \sim \frac{\Gamma \left ( 2\gamma +1 \right )}{2\sigma^{(\gamma)}}\frac{\sin\left ( \pi \gamma  \right )}{\pi}d^{-\varrho}.
\label{power_decay}
\end{equation}

\noindent where $\varrho=1+2\gamma$. Consequently, in the simple case of regular multiplexes with $J=1$, a power-law relation emerge for the transitions between both layers when $d\gg1$ and $N\rightarrow\infty$. Once $0<\gamma<1$, in this process the long-range transitions between different layers decay with exponent $1<\varrho<3$, in a similar way as in the case of fractional diffusion in monolayer regular networks \cite{riascos14,riascos15,michelitsch19}. In Fig.~\ref{ltransicion_gamma_0p5}(a) we show the dependence of $\mathcalboondox{T}_{i(j+N)}^{(\gamma)}$ on $d$ for regular multiplex networks with two layers and $J=1$. As can be observed, the predicted exponent $\varrho$ is in excellent agreement with the results obtained from Eqs.~(\ref{frac_RWLapl}). Besides, Fig.~\ref{ltransicion_gamma_0p5}(a) shows that, for a given $D_x$, the larger the value of $\gamma$, the larger $\mathcalboondox{T}_{i(j+N)}^{(\gamma)}$ is when $d\sim1$ (see inset). On the other hand, in Fig.~\ref{ltransicion_gamma_0p5}(b) we illustrate the dependence of $\mathcalboondox{T}_{i(j+N)}^{(\gamma)}$ on $D_x$. As expected, the larger $D_x$, the smaller (larger) the element of the transition matrix when $d\gg1$ ($d\sim1$). In the case of $d\gg1$, an increase on the interlayer diffusion coefficient also increases $\sigma^{(\gamma)}$, and the latter is inversely proportional to $\mathcalboondox{T}_{i(j+N)}^{(\gamma)}$ [see Eq.~(\ref{frac_RWLapl})]. The results in  Fig.~\ref{ltransicion_gamma_0p5}(b) also confirm that $\varrho$ does not depend on $D_x$ (when $d\gg1$).

Finally, it is worth mentioning that fractional diffusion induces a novel mechanism of interlayer diffusion: fractional node-centric CTRWs are allowed to switch layer and jump to another vertex that may be very far away. For instance, L\'evy RWs in \cite{guo16} are not allowed to switch layer and hop during the same jump. On the other hand, the physical RWs presented in \cite{dedomenico14} reduce to classic RWs on monolayer networks, which are subject to the \textbf{NN} paradigm. However, these fractional node-centric CTRWs exhibit long-range hops on top of monolayer networks (see \cite{riascos14,riascos15,michelitsch19}).


\subsection{Gaussian enhanced diffusion: MSD on regular multiplex networks}
\label{sec_MSD}

\begin{figure}[h!]
\centering
\subfloat[]{
\centering
\includegraphics[width=0.5\linewidth]{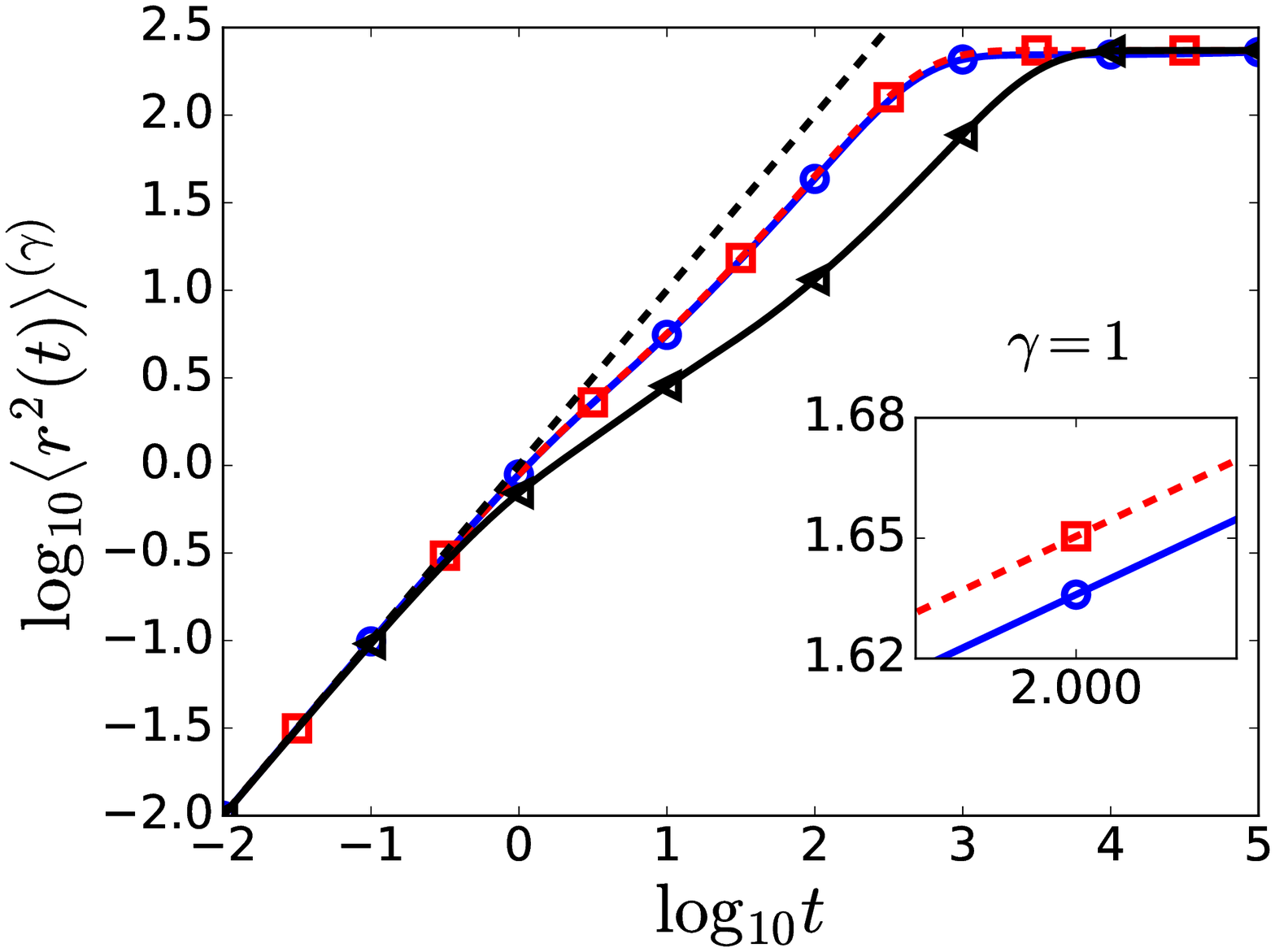}
\label{MSD_comp_a}
}
\subfloat[]{
\centering
\includegraphics[width=0.5\linewidth]{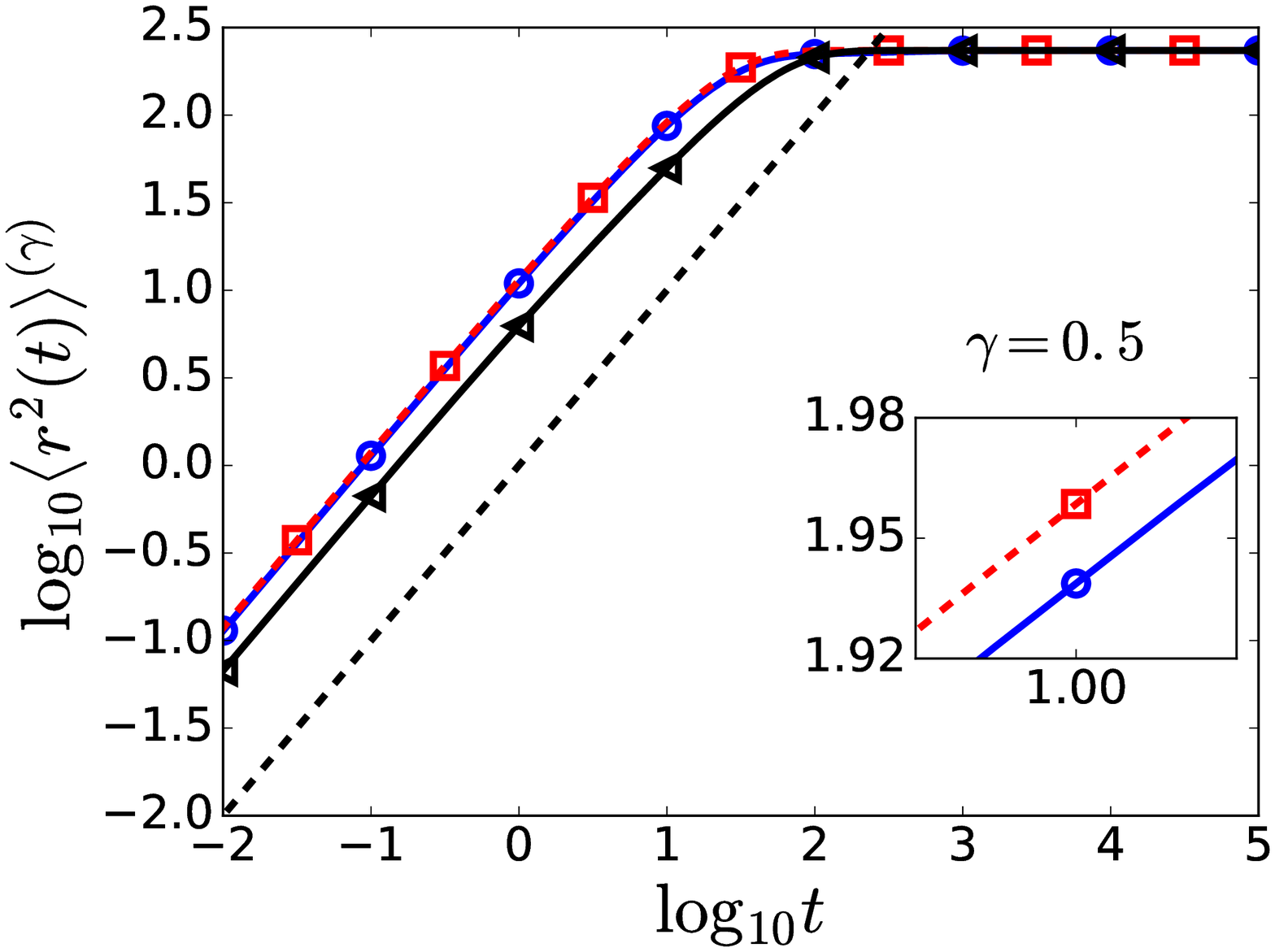}
\label{MSD_comp_b}
}
\caption{Time evolution of $\left \langle r^2(t) \right \rangle^{\left (\gamma \right )}$ on a regular multiplex network with $N=501$ and $J=10$, for $\gamma=1$ (a) and $\gamma=0.5$ (b): $D_x=10^{-4}$ (blue circles), $D_x=A_c/2\approx 0.03$ (red squares), and $D_x=10^{2}$ (black triangles). Red dashed lines indicate the results for the optimal $D_x=A_c/2$. The black dashed line is a guide for the eye to identify a Gaussian behavior. The insets show details of the very small differences between the results for $D_x=10^{-4}$ and $D_x=A_c/2$.}
\label{MSD_comp}
\end{figure}

In this section, we show an example of the nonmonotonic increase in the rate of convergence to the steady state of centric-node CTRWs diffusion. To do so, we study the dependence of the mean square displacement of the walkers, $\left \langle r^2(t) \right \rangle^{(\gamma)}$, on $D_x$ and $\gamma$ [see Eqs.~(\ref{prob_walker_frac}) and (\ref{eq_frac_MSD_multpl})]. In Fig.~\ref{MSD_comp} we show the results for a regular multiplex network without fractional diffusion ($\gamma=1.0$, see left panel) and with it ($\gamma=0.5$, see right panel), for several values of $D_x$: the optimal interlayer coefficient $D_x=A_c/2$, and two aditional values, which are very large and very small in comparison to it. As expected, the results show that, for a given value of $\gamma$, the fastest convergence to the steady state corresponds to the optimal $D_x$. We can observe that the differences between the results for $D_x=A_c/2$ and $D_x\ll A_c/2$ are very small. In both cases the layers are barely coupled due to the very small value of $A_c$ [see Eq.~(\ref{eigen_xi}) when $c=2$]. Nonetheless, when $D_x=A_c/2$, the inter-layer connection is stronger and $\Lambda_2$ reaches a maximum, i.e., the diffusion is enhanced. In the case of very large values of $D_x$, the diffusion of the node-centric CTRWs is hindered once, as $\sigma^{(\gamma)}\approx D_x$, the walkers spend most of their time  switching layers instead of hoping to other nodes inside the layers. On the other hand, for a given value of $D_x$, it can be seen that, the smaller the $\gamma$, the larger $\Lambda_2$, and the faster the diffusion (the diffusion time scale $\tau\sim1/\Lambda_2$). Thus, the previous findings are in good agreement with Eq.~(\ref{algebraic_reg}) and the data in Fig.~\ref{lambda_2_reg}(a).

Finally, it is worth mentioning that the increased algebraic connectivity induced by fractional dynamics is reflected in the long-range navigation of node-centric walkers. As can be observed in Fig.~\ref{MSD_comp}(b), in the case of $\gamma=0.5$, $\left \langle r^2(t) \right \rangle^{(\gamma)}\approx t$ (i.e. $\varepsilon =1$). Thus, a Gaussian behavior emerges from the fractional dynamics in finite circulant multiplex networks with two layers. Other examples of circulant multiplex networks with different values of $N$, $J^1$, $J^2$ and $\gamma$ are presented in the Supplemental Material accompanying this paper, and all of them show perfect agreement with the developed analysis.


\subsection{Optimal diffusion on non-regular multiplexes}
\label{sec_nonreg}

\begin{figure}[h!]
\centering
\subfloat[]{
\centering
\includegraphics[width=0.5\linewidth]{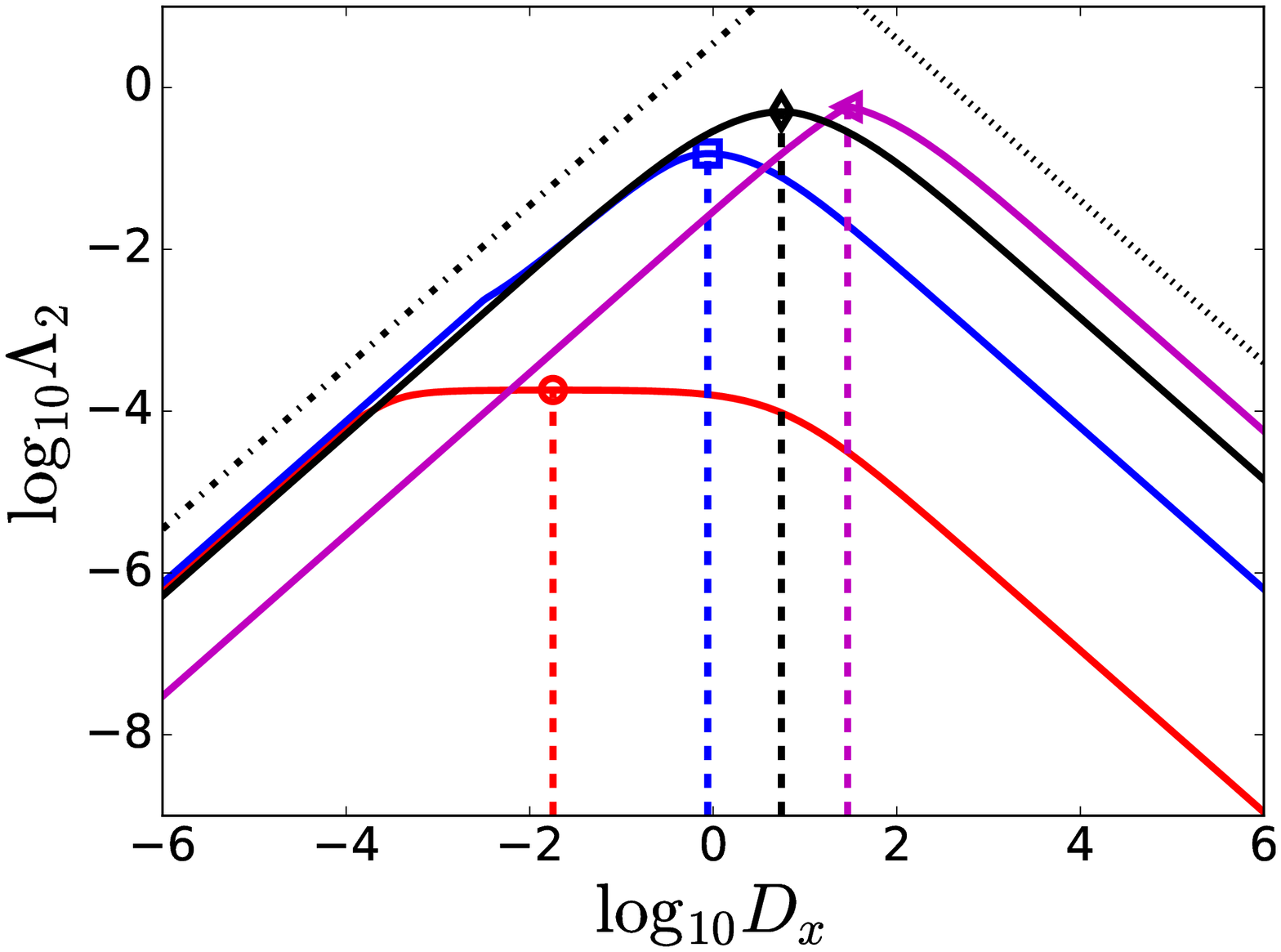}
\label{}
}
\subfloat[]{
\centering
\includegraphics[width=0.5\linewidth]{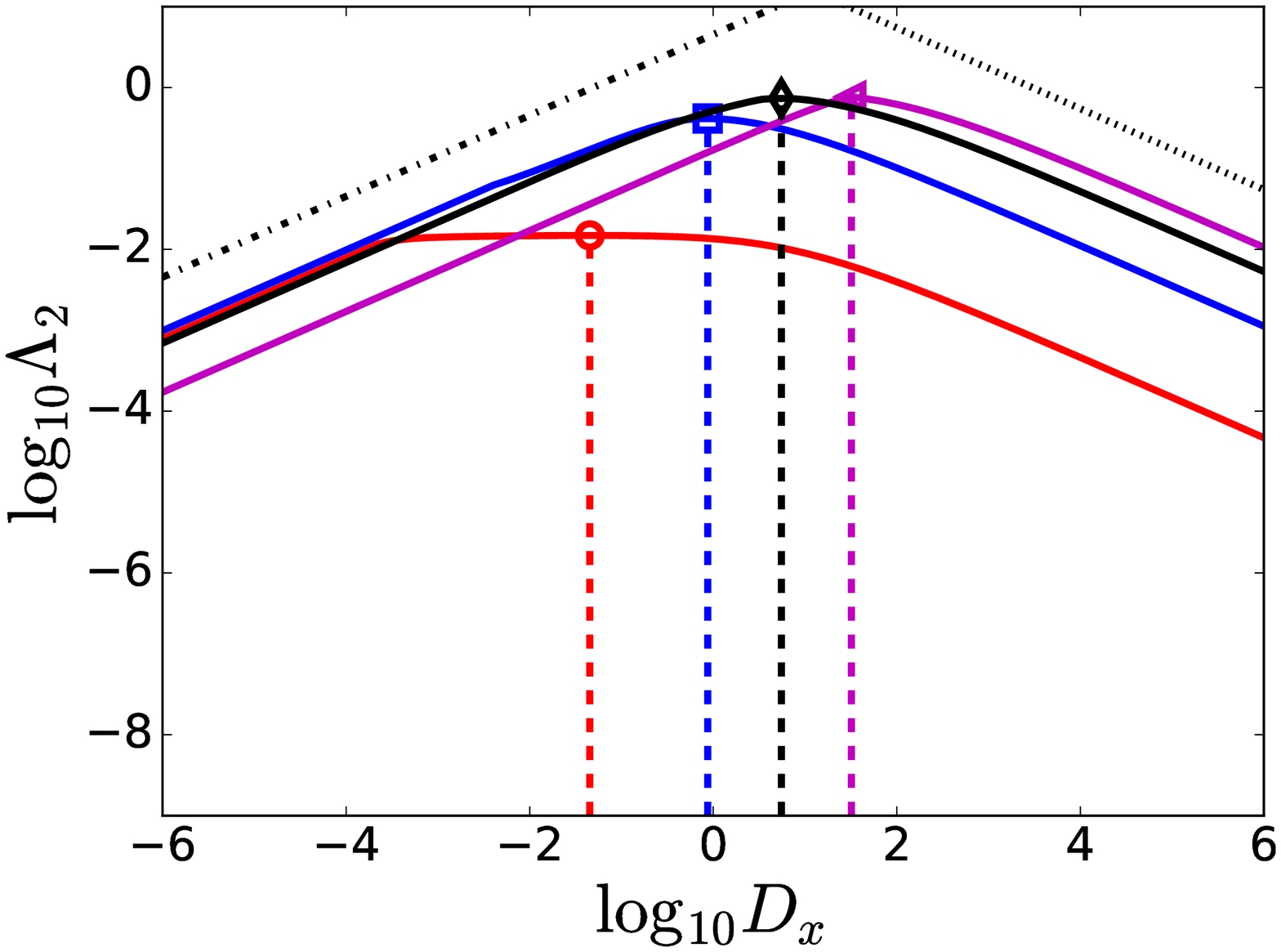}
\label{}
}
\caption{Dependence of $\Lambda_2$ on $D_x$ for multiplex networks with two layers and $N=1001$ nodes, when $\gamma=1$ (a) and $\gamma=0.5$ (b): two circulant layers ($J^1=1$, $J^2=5$, red line), two Bar\'abasi-Albert (BA) networks ($m_1=1$, $m_2=2$, blue line), two Erd\H{o}s-R\'enyi (ER) random graphs ($p_1=0.1$, $p_2=0.05$, black continuous line), and BA-ER layers ($m=1$, $p=0.05$, magenta line) \cite{note1}. Symbols represent the global maximum of $\Lambda_2$ for each series. Black dash-dotted line is a guide for the eye proportional to $D_x^\gamma$, whereas the black dotted line is proportional to $D_x^{-\gamma}$.}
\label{Lambda_2_gamma_1_varias_top}
\end{figure}

In Fig.~\ref{Lambda_2_gamma_1_varias_top} we present examples of multiplexes with circulant and noncirculant layers, when $\gamma=1$ (left panel) and $\gamma=0.5$ (right panel). As can be seen, in all the cases the nonmonotonic trend of $\Lambda_2$ is present. Indeed, when $D_x\rightarrow0$ ($D_x\rightarrow\infty$), its dependence on $D_x$ is similar to $D_x ^{\gamma}$ ($D_x ^{-\gamma}$). For that reason, there is a nonmonotonic increase in the rate of convergence to the steady state of node-centric CTRWs. Besides, in all the topologies tested, the smaller the value of $\gamma$, the larger the global maximum of $\Lambda_2$. Therefore, there exist optimal combinations of $D_x$ and $\gamma$ that enhance diffusion processes of node-centric RWs and make them faster than those obtained when layers are fully coupled and $\gamma=1$. In the case of circulant multiplexes, the findings presented here are in excellent agreement with Eqs.~(\ref{alge_conn_small}) and (\ref{alge_conn_large}). On the other hand, these results suggest that the apparent plateau observed in circulant multiplexes is not present in other topological configurations. It seems that  the more random the layers are, the larger $\Lambda_2$ is. For that reason, finding an optimal value of $D_x$ is more crucial in noncirculant multiplex networks than in circulant ones.

\section{Conclusions}
\label{conclusions}

In this work, we have extended the continuous time fractional diffusion framework (for simple networks) introduced in Refs. \cite{riascos14,riascos15,michelitsch19} to multiplex networks with undirected and unsigned layers. Hence fractional diffusion is defined here in terms of the fractional supra-Laplacian matrix of the system, i.e., the combinatorial supra-Laplacian matrix of the multiplex $\mathbf{\mathcal{L}}^\mathcal{M}$ to a power $\gamma$, where $0<\gamma<1$. For the purpose of deriving exact analytical expressions, we have considered only diffusion processes in which $\mathbf{\mathcal{L}}^\mathcal{M}$ is a symmetric matrix.

We have focused our discussion on the characterization of Poissonian node-centric continuous-time random-walks on circulant multiplexes with two layers, and explored the combined effect of inter and intra layer diffusion with fractional dynamics. We have directed our attention to (i) the effect of the fractional dynamics on the nonmonotonic increase in the rate of convergence to the steady state of such process, (ii) the existence of an optimal regime that depends on both the inter-layer coupling $D_x$ and on the fractional parameter $\gamma$, and (iii) the emergence of a new type of long-range navigation on multiplex networks. For circulant multiplexes, analytical expressions were obtained for the main quantities involved in these dynamics, namely: the eigenvalues and eigenvectors of the combinatorial supra-Laplacian matrix and of the normalized supra-Laplacian matrix, the fractional strength of the nodes, the fractional transition matrix, the probability of finding the walkers at time $t$ on any node of a given layer, and the mean-square displacement for fractional dynamics. For other multiplex topologies some of these quantities were obtained by numerical evaluations.

We have shown that, for a given circulant multiplex network, the more intense the fractional diffusion (i.e., the smaller the paramenter $\gamma$), the larger the algebraic connectivity of the normalized supra-Laplacian matrix, denoted as $\Lambda_2$. Since the diffusion time scale of the Poissonian node-centric CTRWs on the multiplex $\tau$ is inversely proportional to $\Lambda_2$, the smaller the value of $\gamma$, the faster the convergence to the steady state is (i.e., the smaller $\tau$ is). Additionally, in multiplexes with two layers, both $\Lambda_2$ and $\tau$ exhibit a nonmonotonic dependence on $D_x$, respectively, whether or not there are fractional diffusion. Consequently, the rate of convergence to the steady state must be optimized when using fractional dynamics.

On the other hand, in the simple case of circulant (regular) multiplexes with $J=1$, we have illustrated that, once the fractional diffusion is present (i.e., $0<\gamma<1$), long-range transitions between different layers appear. Indeed, a new continuous-time random walk process appears, since here walkers are allowed to (i) switch layer and (ii) perform long hops to another distant vertex in the same jump. Additionally, in the thermodinamic limit, we have shown that the probability of long range transitions decay according to a power-law with exponent $1<\varrho<3$, in a similar way as in the case of fractional diffusion in monolayer regular networks \cite{riascos14,riascos15,michelitsch19}. We have also shown that the larger $D_x$, the smaller (larger) the transition probability between nodes that are very far away (very close).

Finally, the evaluation of $\equiv \left \langle r^2(t) \right \rangle$ indicates the existence of the optimal regime that depends on both $D_x$ and the fractional parameter $\gamma$. On the other hand, we have shown that the enhaced diffusion induced by fractional dynamics on finite circulant multiplexes exhibits a Gaussian behavior ($\left \langle r^2(t) \right \rangle\sim t$) before saturation appears.

The introduction of fractional dynamics on multiplex networks opens new possibilities for analyzing and optimizing (anomalous) diffusion on such arrangements. For instance, given the attention devoted recently to the optimal diffusion dynamics on directed multiplex networks, the generalization of this framework to such systems can be of great interest. On the other hand, the emerging long-range transitions can enhance the efficiency of the navigation on non-circulant multiplex topologies. While studing the dependence of $\Lambda_2$ on $D_x$ in circulant multiplexes, we have found an apparent plateau in $\Lambda_2$ that is not present in other topological configurations. Indeed, it seems that  the more random the layers are, the more pronounced is the optimal $\Lambda_2$. Thus, finding optimal combinations of $D_x$ and $\gamma$ seems more crucial in noncirculant multiplex networks than in circulant ones. For that reason, an exhaustive research on the dependence of $\Lambda_2$ on noncirculant topologies with fractional diffusion is being conducted and it will be published elsewhere.

\begin{acknowledgments}

This work was supported by the Brazilian agencies CAPES and CNPq through the Grants 151466/2018-1 (AA-P) and 305060/2015-5 (RFSA). RFSA also acknowledges the support of the National Institute of Science and Technology for Complex Systems (INCT-SC Brazil).

\end{acknowledgments}


\section*{APPENDIX A: Eigenvalues and Eigenvectors of the combinatorial supra-Laplacian matrix for circulant multiplexes with two layers.}
\label{Eigen_Comb_Supr_L_App}

Let $\mathcal{M}$ be an undirected multiplex network with $N$ nodes and $M=2$ layers, all of which consist of interacting cycle graphs. According to Eqs.~(\ref{comb_circ_ini1}) and (\ref{comb_circ_ini2}), the eigenvalues of $\mathbf{\mathcal{L}}^\mathcal{M}$, denoted as $\mu_g$ for $g\in\left \{ 1,\cdots,2N \right \}$, meet the following condition:

\begin{equation}
\mathrm{det}\left ( \Upsilon  \right ) =\mathrm{det}\left ( \begin{matrix}\mathrm{\Xi}^1 -\mu_g\mathbf{I}_N & -D_x\mathbf{I}_N\\
-D_x\mathbf{I}_N & \mathrm{\Xi}^2 -\mu_g\mathbf{I}_N
\end{matrix} \right )=0.
\label{comb_circ_ini3}
\end{equation}

\noindent where $\mathrm{det}\left (  \mathbf{X} \right )$ refers to the determinant of a matrix $\mathbf{X}$. Taking into account that (i) the blocks of $\Upsilon$ are square matrices of the same order, and (ii) matrices $-D_x\mathbf{I}_N$ and $\mathrm{\Xi}^2 -\mu_g\mathbf{I}_N$ commute, it is possible to  show \cite{Silvester00} that Eq.~(\ref{comb_circ_ini3}) reduces to

\begin{equation}
\mathrm{det}\left (   (\mathrm{\Xi}^1 -\mu_g\mathbf{I}_N)(\mathrm{\Xi}^2 -\mu_g\mathbf{I}_N)-(-D_x\mathbf{I}_N)(-D_x\mathbf{I}_N)\right )=0.
\label{comb_circ_ini4}
\end{equation}

\noindent By definition, $\mathrm{\Xi}^\alpha=\mathrm{diag}\left (  \xi_1^\alpha ,\cdots, \xi_{N}^\alpha  \right )$, where $\xi_m^\alpha$ are the eigenvalues of $\mathbf{C}^\alpha$ [see Eqs.~(\ref{comb_circ_ini1}) and (\ref{eigen_xi})], and $m\in\left \{ 1,\cdots,N \right \}$. Consequently, Eq.~(\ref{comb_circ_ini4}) is equivalent to the following $N$ equations:

\begin{equation}
(\mu_g)^2+\mu_g(\xi_m^1+\xi_m^2)+(\xi_m^1\xi_m^2-D_x^2)=0.
\label{comb_circ_ini5}
\end{equation}

\noindent For a given value of $m$, we donote the two roots of Eq.~(\ref{comb_circ_ini5}) as $\mu_{2m-1}$ and $\mu_{2m}$, respectively. Thus, we obtain Eqs.~(\ref{lapl_eigenA}) and (\ref{lapl_eigenB}).

On the other hand, the corresponding eigenvector of $\mu_g$, denoted by $\vec{\iota}_g$, can be calculated from

\begin{equation}
\left ( \begin{matrix}\mathrm{\Xi}^1 -\mu_g\mathbf{I}_N & -D_x\mathbf{I}_N\\
-D_x\mathbf{I}_N & \mathrm{\Xi}^2 -\mu_g\mathbf{I}_N
\end{matrix} \right )\vec{\iota}_g=
\left ( \begin{matrix}\mathrm{\Xi}^1 -\mu_g\mathbf{I}_N & -D_x\mathbf{I}_N\\
-D_x\mathbf{I}_N & \mathrm{\Xi}^2 -\mu_g\mathbf{I}_N
\end{matrix} \right )\left ( \begin{matrix}
\vec{v}_g^1\\
\vec{v}_g^2
\end{matrix} \right )=0,
\label{comb_circ_ini6}
\end{equation}

\noindent where $\vec{v}_g^1$ and $\vec{v}_g^2$ are $N\times 1$ vectors. According to Eq.~(\ref{comb_circ_ini6}), the elements of $\vec{v}_g^1$ and of $\vec{v}_g^2$ should meet simultaneously the following conditions:

\begin{equation}
\left\{\begin{matrix}
\frac{\xi_m^1-\mu_g}{D_x}\left ( \vec{v}_g^1 \right )_m=\left ( \vec{v}_g^2 \right )_m\\
\frac{\xi_m^2-\mu_g}{D_x}\left ( \vec{v}_g^2 \right )_m=\left ( \vec{v}_g^1 \right )_m
\end{matrix}\right..
\label{comb_circ_ini7}
\end{equation}

\noindent It is possible to see that the previous restrictions are equivalent to Eq.~(\ref{comb_circ_ini5}). Therefore, in the case of $g\in\left \{ 2m-1,\,2m \right \}$, Eq.~(\ref{comb_circ_ini7}) requires that only the elements $\left ( \vec{v}_g^1 \right )_m$ and $\left ( \vec{v}_g^1 \right )_m$ are non-zero. Consequently, to normalize $\vec{\iota}_g$, we set $\left ( \vec{v}_g^1 \right )_m=1/T_g$ and $\left ( \vec{v}_g^2 \right )_m=(\xi_m^1-\mu_g)/(D_xT_g)$, where

\begin{equation}
T_g = \sqrt{1+\left ( \frac{\xi_m^1-\mu_g}{D_x} \right )^2}=\sqrt{1+\left ( M_g \right )^2},
\label{comb_circ_ini9}
\end{equation}

\noindent for $g\in\left \{ 2m-1,\,2m \right \}$ [see Eq.~(\ref{coeff_M_g})]. According to the previous results, given the matrix definedby the right hand side of Eq.~(\ref{comb_circ_ini2}) and its corresponding eigenvectors $\vec{\iota}_g$, the matrix $\mathbf{\Omega}=\left ( \begin{matrix} \vec{\iota}_1 &\cdots  & \vec{\iota}_{2N}\end{matrix} \right )$ has elements

\begin{equation}
\left.\begin{matrix}
\mathbf{\Omega}_{fg}=1/T_g\\
\mathbf{\Omega}_{(f+N)g}=M_g/T_g\
\end{matrix}\right\},
\label{comb_circ_ini9_2}
\end{equation}

\noindent for $g\in\left \{ 1,\cdots,2N \right \}$ and $f=1+\left \lfloor (g-1)/2 \right \rfloor$ (i.e., $f\in\left \{ 1,\cdots,N \right \}$), and zero otherwise.

Finally, the eigenvectors of the combinatorial supra-Laplacian matrix $\mathbf{\mathcal{L}}^\mathcal{M}$, i.e. $\vec{\psi}_g $, and the matrix $\mathbf{Q}=\left ( \begin{matrix} \vec{\psi}_1 &\cdots  & \vec{\psi}_{2N}\end{matrix} \right )$ [in Eqs.~(\ref{Q_layer1}) and (\ref{Q_layer2})] can be obtained from

\begin{equation}
\mathbf{Q}=\mathbf{F}\mathbf{\Omega},
\label{comb_circ_ini10}
\end{equation}

\noindent as can be seen from Eq.~(\ref{def_F}).


\section*{APPENDIX B: Eigenvalues and Eigenvectors of the normalizedl supra-Laplacian matrix for circulant multiplexes with two layers.}
\label{Eigen_Norm_Supr_L_App}

Let $\mathcal{M}$ be an undirected $N-$node multiplex network in which all its $M=2$ layers consist of interacting cycle graphs. According to Eqs.~(\ref{frac_degree})-(\ref{frac_RWLapl}), the normalized fractional supra-Laplacian matrix of the multiplex $\mathcalboondox{L}^{\left (\gamma \right )}$ is given by $\mathcalboondox{L}^{\left (\gamma \right )}=\mathcal{K}^{-1}\left ( \mathbf{\mathcal{L}}^\mathcal{M} \right )^\gamma$, where

\begin{equation}
\mathcal{K}^{-1}=\left ( \begin{matrix}\frac{1}{\sigma_1^{\left (\gamma \right )}}\mathbf{I}_N & \mathbf{0}\\
\mathbf{0} & \frac{1}{\sigma_2^{\left (\gamma \right )}}\mathbf{I}_N
\end{matrix} \right ).
\label{comb_circ2_ini1}
\end{equation}

\noindent Taking into account Eqs.~(\ref{frac_Lapl}), (\ref{def_F}), (\ref{comb_circ_ini9_2}) and (\ref{comb_circ_ini10}), it is possible to write

\begin{align}
\mathbf{F}^{-1} \mathcalboondox{L}^{\left (\gamma \right )}\mathbf{F}&=\mathbf{F}^{-1} \mathcal{K}^{-1} \mathbf{F}\mathbf{\Omega} \mathbf{\Delta}^\gamma \mathbf{\Omega}^{-1} \mathbf{F}^{-1} \mathbf{F}\nonumber \\
&=\mathcal{K}^{-1}\mathbf{\Omega} \mathbf{\Delta}^\gamma \mathbf{\Omega}^{T},
\label{comb_circ2_ini2}
\end{align}

\noindent where $\mathbf{\Omega}^{-1} = \mathbf{\Omega}^{T}$, since $\mathbf{Q}=\mathbf{F}\mathbf{\Omega}$, $\mathbf{Q}\mathbf{Q}^\dagger=\mathbf{I}_{2N}$, and $\mathbf{I}_{2N}$ represents the $2N \times 2N$ identity matrix. Thus, the eigenspectrum of $\mathcalboondox{L}^{\left (\gamma \right )}$ is equal to that of $\mathcal{K}^{-1}\mathbf{\Omega} \mathbf{\Delta}^\gamma \mathbf{\Omega}^{T}$.

Considering that $\mathbf{\Delta}^\gamma= \mathrm{diag}\left ( \mu_1^\gamma,\mu_2^\gamma,\cdots,\mu_{2N}^\gamma \right )$, as well as the definition of $\Omega$ [Eq.~(\ref{comb_circ_ini9_2})], the right hand side of Eq.~(\ref{comb_circ2_ini2}) can be rewritten as

\begin{equation}
\mathcal{K}^{-1}\mathbf{\Omega} \mathbf{\Delta}^\gamma \mathbf{\Omega}^{T}=\left ( \begin{matrix}
\frac{1}{\sigma_1^{(\gamma )}}\mathbf{D}^1 & \frac{1}{\sigma_1^{(\gamma )}}\mathbf{D}^3\\
\frac{1}{\sigma_2^{(\gamma )}}\mathbf{D}^3 & \frac{1}{\sigma_2^{(\gamma )}}\mathbf{D}^2
\end{matrix} \right ),
\label{comb_circ2_ini3}
\end{equation}

\noindent where $\mathbf{D}^{1}$, $\mathbf{D}^{2}$ and $\mathbf{D}^{3}$ are $N\times N$ diagonal matrices, whose respective $m$ elements are given by

\begin{equation}
\left ( \mathbf{D}^1 \right )_m=\frac{\mu_{2m-1}^\gamma }{1+(S+C)^2}+\frac{\mu_{2m}^\gamma }{1+(S-C)^2},
\label{comb_circ2_ini4}
\end{equation}

\begin{equation}
\left ( \mathbf{D}^2 \right )_m=\frac{\mu_{2m-1}^\gamma (S+C)^2}{1+(S+C)^2}+\frac{\mu_{2m}^\gamma (S-C)^2}{1+(S-C)^2},
\label{comb_circ2_ini5}
\end{equation}

\noindent and

\begin{equation}
\left ( \mathbf{D}^3 \right )_m=\frac{\mu_{2m-1}^\gamma (S+C)}{1+(S+C)^2}+\frac{\mu_{2m}^\gamma (S-C)}{1+(S-C)^2},
\label{comb_circ2_ini6}
\end{equation}

\noindent for $m\in\left \{ 1,\cdots,N \right \}$, where

\begin{align}
S=\frac{\xi_m^1-\xi_m^2}{2D_x}=\frac{A_m^1-A_m^2}{2D_x},
\end{align}

\noindent and $C^2=1+S^2$. Notice that, by making use of Eqs.~(\ref{coeff_M_g}) and (\ref{comb_circ_ini9}), it is possible to express some terms in Eqs.~(\ref{comb_circ2_ini4})-(\ref{comb_circ2_ini6}) in terms of $M_g$ and $T_g$. Being more specific, we can write $\left ( M_{2m-1} \right )^2=(S+C)^2$ and $\left ( M_{2m} \right )^2=(S-C)^2$, as well as $\left ( T_{2m-1} \right )^2=1+(S+C)^2$ and $\left ( T_{2m} \right )^2=1+(S-C)^2$ [see also the definition of $\Omega$ given by Eq.~(\ref{comb_circ_ini9_2})].

According to Eq.~(\ref{comb_circ2_ini3}), to calculate the eigenvalues of $\mathcalboondox{L}^{\left (\gamma \right )}$, denoted as $\lambda_g$ for $g\in\left \{ 1,\cdots,2N \right \}$, we solve the following equation:

\begin{equation}
\mathrm{det}\left ( \Phi  \right )=\mathrm{det}\left ( \begin{matrix}
\frac{1}{\sigma_1^{(\gamma )}}\mathbf{D}^1 -\lambda_g \mathbf{I}_N& \frac{1}{\sigma_1^{(\gamma )}}\mathbf{D}^3\\
\frac{1}{\sigma_2^{(\gamma )}}\mathbf{D}^3 & \frac{1}{\sigma_2^{(\gamma )}}\mathbf{D}^2 -\lambda_g \mathbf{I}_N
\end{matrix} \right )=0.
\label{comb_circ2_ini7}
\end{equation}

\noindent Since (i) the blocks of $\Phi$ are square matrices of the same order, and (ii) matrices $\frac{1}{\sigma_2^{(\gamma )}}\mathbf{D}^3$ and $\frac{1}{\sigma_2^{(\gamma )}}\mathbf{D}^2 -\lambda_g \mathbf{I}_N$ commute, it is possible to show \cite{Silvester00} that Eq.~(\ref{comb_circ2_ini7}) reduces to

\begin{equation}
\mathrm{det}\left (   \left ( \frac{1}{\sigma_1^{(\gamma )}}\mathbf{D}^1 -\lambda_g \right )\left ( \frac{1}{\sigma_2^{(\gamma )}}\mathbf{D}^2 -\lambda_g \mathbf{I}_N \right )-\left (  \frac{1}{\sigma_1^{(\gamma )}}\mathbf{D}^3\right )\left ( \frac{1}{\sigma_2^{(\gamma )}}\mathbf{D}^3 \right )\right )=0.
\label{comb_circ2_ini8}
\end{equation}

\noindent Finally, the eigenspectrum of $\mathcalboondox{L}^{\left (\gamma \right )}$ is obtained by calculating the roots of the following $N$ equations:

\begin{equation}
(\lambda_g)^2+\lambda_g\left ( \frac{1}{\sigma_1^{(\gamma )}}\left ( \mathbf{D}^1 \right )_m+\frac{1}{\sigma_2^{(\gamma )}}\left ( \mathbf{D}^2 \right )_m \right )+\left ( \frac{1}{\sigma_1^{(\gamma )}\sigma_2^{(\gamma )}}\left ( \mathbf{D}^1 \right )_m\left ( \mathbf{D}^2 \right )_m - \frac{1}{\sigma_1^{(\gamma )}\sigma_2^{(\gamma )}} \left ( \left ( \mathbf{D}^3 \right )_m \right )^2 \right )=0,
\label{comb_circ2_ini9}
\end{equation}

\noindent which are equivalent to Eq.~(\ref{comb_circ2_ini8}). For a given value of $m\in\left \{ 1,\cdots,N \right \}$, we denote the two roots of Eq.~(\ref{comb_circ2_ini9}) as $\lambda_{2m-1}$ and $\lambda_{2m}$, respectively. Thus, we obtain obtain Eqs.(\ref{RWlapl_eigenA})-(\ref{nu_def}).

\section*{APPENDIX C: Transitions between nodes that are located in different layers.}
\label{Transitions_finite}

Let $\mathcal{M}$ be an undirected multiplex network with $M=2$ layers, both of which are cycle graphs (i.e., $J=1$). Let us consider two nodes in $\mathcal{M}$, which are located in different layers: $i$ is at layer 1 and $j$ is at layer 2. Following Refs. \cite{riascos14,riascos15,michelitsch19}, by using Eqs.~(\ref{Q_layer1})-(\ref{coeff_M_g})  and conducting the necessary manipulation in Eq.~(\ref{element_frac_Lapl}), the element of the fractional supra-Laplacian that refers to $i$ and $j$ can be approximated as follows:

\begin{align}
\left ( \mathbf{\mathcal{L}}^\mathcal{M} \right )_{i(j+N)}^\gamma &=\sum _{m=1}^{N} -\frac{1}{2N}\left ( A_m+2D_x \right )^\gamma  \exp\left ( \mathfrak{i}\frac{2\pi }{N}\left ( j-i \right ) \left ( m-1 \right )\right )\nonumber\\
&+\sum _{m=1}^{N} \frac{1}{2N}A_m^\gamma  \exp\left ( \mathfrak{i}\frac{2\pi }{N}\left ( j-i \right ) \left ( m-1 \right )\right )\nonumber\\
&= \frac{1}{2N}\sum _{m=1}^{N}\left (A_m^\gamma -\left ( A_m+2D_x \right )^\gamma  \right ) \exp\left ( \mathfrak{i}\theta_{m}d\right )\nonumber\\
&\approx \frac{1}{2N} \sum _{m=1}^{N} \left ( A_m^\gamma -\gamma A_m \left (2D_x \right )^{\gamma-1} \right )\exp\left ( \mathfrak{i}\theta_{m}d\right )-\frac{1}{2N}\sum _{m=1}^{N} \left ( 2D_x \right )^\gamma \exp\left ( \mathfrak{i}\theta_{m}d\right )
\label{element_frac_Lapl_regular_app}
\end{align}

\noindent for $\left | D_x \right |>\left | A_m \right |$ and $D_x>0$, where $d\equiv d_{i^1\rightarrow j^1}$, $\theta_m=2\pi\left ( m-1 \right )/N$, and $A_m=2+2\cos\left ( 2\theta_m\right )$ [see Eq.(\ref{eigen_xi}) for $J^\alpha=1$ and $\alpha \in \left \{ 1,2 \right \}$]. If $D_x=0$, then $A_m^\gamma -\left ( A_m+2D_x \right )^\gamma = 0$, and this element of the fractional supra-Laplacian is equal to zero.

In the limit $N\rightarrow\infty$, the sums in Eq.~(\ref{element_frac_Lapl_regular_app}) can be replaced by integrals, so that

\begin{widetext}

\begin{align}
\left ( \mathbf{\mathcal{L}}^\mathcal{M} \right )_{i(j+N)}^\gamma &=\frac{1}{2}\left ( \frac{1}{2\pi }\int _0^{2\pi }A_m^\gamma \exp(\mathfrak{i}d\theta_m)\mathrm{d\theta}_m\right )
-\frac{\gamma \left ( 2Dx \right )^ {\gamma-1} }{2}\lim_{\zeta \rightarrow 1}\left ( \frac{1}{2\pi }\int _0^{2\pi }A_m^\zeta \exp(\mathfrak{i}d\theta_m)\mathrm{d\theta}_m\right )\nonumber\\
&-\frac{\left ( 2Dx \right )^ {\gamma}}{2}\left ( \frac{1}{2\pi }\int _0^{2\pi } \exp(\mathfrak{i}d\theta_m)\mathrm{d\theta }_m\right ).
\label{sum_integrals}
\end{align}

\end{widetext}

\noindent Taking into account that (i) the last right-hand term in Eq.~(\ref{sum_integrals}) is equal to $-\left ( 2Dx \right )^ {\gamma}/2$ for $d=0$ and 0 otherwise, and (ii) the analytical results

\begin{align}
\frac{1}{2\pi }\int _0^{2\pi }A_m^\gamma \exp(\mathfrak{i}d\theta_m)\mathrm{d\theta }=-\frac{\Gamma \left ( d-\gamma  \right )\Gamma \left ( 2\gamma +1 \right )}{\pi \Gamma \left ( 1+\gamma +d \right )}\sin\left ( \pi \gamma  \right ),
\label{integral}
\end{align}

\noindent and

\begin{align}
\lim_{\gamma \rightarrow 1}\frac{\Gamma \left ( d-\gamma  \right )\Gamma \left ( 2\gamma +1 \right )}{\pi \Gamma \left ( 1+\gamma +d \right )}\sin\left ( \pi \gamma  \right )=K_d,
\label{lim_integral}
\end{align}

\noindent we obtain Eq.(\ref{element_frac_Lapl_regular_inf}) [see \cite{riascos14,riascos15,michelitsch19,Zoia07} for further details about the derivation of Eqs.~(\ref{integral}) and (\ref{lim_integral})].

Finally, note that Eq.(\ref{frac_degree_regJ1_inf}) can be derived from Eq.~(\ref{frac_degree_reg}) by using the same process presented in this Appendix and considering $d=0$ ($\left ( \mathbf{\mathcal{L}}^\mathcal{M} \right )_{ii}^\gamma = \sigma^{(\gamma)}$).




\begin{thebibliography}{99}
\bibitem{note1}
In the BA-random model, $m$ represents the number of edges to attach from a new node to existing nodes.
In the ER-model, $p$ is the probability for edge creation between two nodes.
\end{thebibliography}
\end{document}